\documentclass[aps,prl,twocolumn,10pt, amsmath,amssymb,longbibliography,superscriptaddress,floatfix,nobibnotes]{revtex4-2}

\usepackage{graphicx}
\usepackage{dcolumn}
\usepackage{bm}
\usepackage{amsmath}
\usepackage{amssymb}
\usepackage{booktabs}
\usepackage{multirow}
\usepackage{tabularx}
\usepackage{siunitx}
\usepackage[colorlinks=true,citecolor=blue,linkcolor=blue]{hyperref}

\graphicspath{{./Figures/}}

\begin{document}

\title{Correlated interlayer quantum Hall state in large-angle twisted trilayer graphene}

\author{Dohun Kim}
\thanks{These authors contributed equally.}
\affiliation{Department of Physics and Chemistry, Daegu Gyeongbuk Institute of Science and Technology (DGIST), Daegu 42988, Republic of Korea}

\author{Gyeoul Lee}
\thanks{These authors contributed equally.}
\affiliation{Department of Physics, Ulsan National Institute of Science and Technology, Ulsan 44919, 
Republic of Korea}

\author{Nicolas Leconte}
\affiliation{Department of Physics, University of Seoul, Seoul 02504, Korea}

\author{Seyoung Jin}
\affiliation{Department of Physics, Pohang University of Science and Technology, Pohang, 37673, Republic of Korea}

\author{Takashi Taniguchi}
\affiliation{Research Center for Materials Nanoarchitectonics, National Institute for Materials Science, 1-1 Namiki, Tsukuba 305-0044, Japan}

\author{Kenji Watanabe}
\affiliation{Research Center for Electronic and Optical Materials, National Institute for Materials Science, 1-1 Namiki, Tsukuba 305-0044, Japan}

\author{Jeil Jung}
\affiliation{Department of Physics, University of Seoul, Seoul 02504, Korea}

\author{Gil Young Cho}
\thanks{gilyoungcho@kaist.ac.kr}
\affiliation{Department of Physics, Korea Advanced Institute of Science and Technology, Daejeon 34141, Republic of Korea}
\affiliation{Center for Artificial Low Dimensional Electronic Systems, Institute for Basic Science, Pohang 37673, Korea}

\author{Youngwook Kim}
\thanks{y.kim@dgist.ac.kr }
\affiliation{Department of Physics and Chemistry, Daegu Gyeongbuk Institute of Science and Technology (DGIST), Daegu 42988, Republic of Korea}

\date{\today} 

\begin{abstract}
\noindent
ABSTRACT: Trilayer graphene offers systematic control of its electronic structure through stacking sequence and twist geometry, providing a versatile platform for correlated states. Here we report magnetotransport in large-angle twisted trilayer graphene with a twist angle of about $5^\circ$. The data reveal an electron–hole asymmetry that can be captured by introducing layer-dependent potential shifts. At charge neutrality ($\nu_{\rm tot}=0$), three low-resistance states appear, which Hartree–Fock mean-field analysis attributes to spin-resolved helical edge modes in the quantum Hall regime, analogous to quantum spin Hall–like configurations. At $\nu_{\rm tot}=-1$, we also observe suppressed resistance when the middle and bottom layers are each half filled while the top layer remains inert at $\nu=-2$, consistent with an interlayer excitonic phase in the quantum Hall regime. These results demonstrate correlated interlayer quantum Hall phases in large-angle twisted trilayer graphene, combining spin-resolved helical edge transport with excitonic order.

\end{abstract}

\maketitle
\begin{figure*}
\begin{center}
\includegraphics{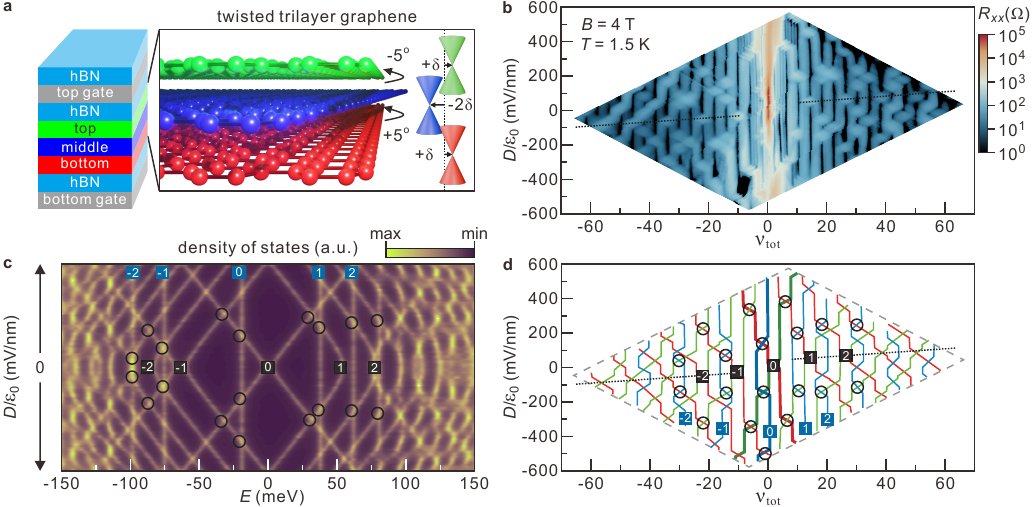}
\caption{{\bf Landau level crossings and quantum Hall states in large-angle twisted trilayer graphene.} {\bf (a)} Schematic of the TTG with a $5^\circ$ twist heterostructure and the layer-resolved potential profile across the graphene layers at zero displacement field.{\bf (b)} Color rendition of the longitudinal resistance $R_{xx}$ in the ($\nu_{\rm tot}$, $D/\varepsilon_{0}$) plane at $T = 1.5$~K and $B = 4$~T. {\bf (c)} Calculated density of states in the $E$–$D/\varepsilon_{0}$ plane at $B = 4$~T. At $D = 0$, numbers in black boxes mark Landau level indices of the top and bottom layers, while numbers in blue boxes mark those of the middle layer. Black circles indicate Landau level crossings. {\bf (d)} Schematic representation of (b). The green, blue, and red curves denote Landau levels of the top, middle, and bottom layers, respectively. Thicker curves indicate the lowest Landau levels, while thinner curves correspond to higher-index levels. Numbers in black boxes show the Landau level index $N$, and black circles indicate crossing points. The dashed line highlights the condition where Landau levels of the top and bottom layers have the same index.}

\label{Fig1}
\end{center}
\end{figure*}

Multilayer graphene exhibits stacking-dependent electronic structures and transport. In the trilayer case, Bernal (ABA) stacking preserves mirror symmetry about the middle layer, yielding coexisting monolayer- and bilayer-like bands with semimetallic overlap \cite{Koshino2011, Taychatanapat2011, Henriksen2012PRX, Serbyn2013, Lee2013, Campos2016, Stepanov2016, Datta2017, Zibrov2018, Che2020, Chen2024}. Rhombohedral (ABC) stacking instead produces cubic dispersion with an enhanced low-energy density of states, where a perpendicular displacement field polarizes the outer layers and opens a sizable gap \cite{Koshino2009, Zhang2010, Lui2011, Jung2013, Jia2013, Lee2014, Shi2020, Zhou2021HQM, Zhou2021SC, Winterer2024}. Beyond these natural stackings, introducing a relative twist between layers reconstructs the electronic spectrum and enables a new class of correlated and topological phases. In the low-angle regime, including magic-angle bilayer, trilayer, and quadrilayer graphene, flat bands host correlated insulating phases, superconductivity,  anomalous Hall states, and other symmetry-broken orders \cite{Cao2018Ins, Cao2018SC, Sharpe2019, Serlin2020, Xie2021FCI, Cao2021Nematic, Park2021TTG, Hao2021ATTG, Park2022MAMLG, Burg2022ATQG}. Such sensitivity to stacking and twist provides the foundation for exploring correlated phases.

 Once the twist angle exceeds the magic angle, the increasing interlayer misalignment produces a substantial momentum mismatch that effectively decouples the electron systems, while interlayer interactions remain significant owing to the atomic interlayer separation. This makes it an ideal ground to realize quantum Hall systems with interlayer excitonic condensation, yet the layers themselves remain independently gate-tunable.  For instance, twisted bilayer and twisted double bilayer graphene have provided model systems where excitonic quantum Hall states, possessing gaps significantly larger than those in semiconductor analogues, can be stabilized without the need for insulating barriers~\cite{Kim2019, Kim2021, Shi2022WSe2EC, Kim2023, Kim2023AdvSci, Kim2025, Li2024}.

Motivated by the progress described above, here we investigate quantum Hall states in twisted trilayer graphene (TTG) with a large twist angle close to $5^\circ$. Although trilayer graphene has been extensively studied \cite{Koshino2011, Taychatanapat2011, Henriksen2012PRX, Serbyn2013, Lee2013, Campos2016, Stepanov2016, Datta2017, Zibrov2018, Che2020, Chen2024,Koshino2009, Zhang2010, Lui2011, Jung2013, Jia2013, Lee2014, Shi2020, Zhou2021HQM, Zhou2021SC, Winterer2024,Cao2021Nematic, Park2021TTG}, the quantizing regime at large twist angles remains unexplored, even though the additional layer degree of freedom is expected to greatly enrich the landscape of emerging correlated phases. Large-angle TTG composed of three decoupled Dirac systems allows access to layer-resolved filling sequences that cannot be realized in bilayers. Indeed, in this regime we uncover an asymmetry between electron and hole doping, along with a series of unconventional low-resistance states at charge neutrality. Mean-field Hartree–Fock analysis indicates that these states are insulating phases that host spin-resolved helical edge configurations in the quantum Hall regime, accompanied by suppressed $R_{xy}$, rather than fully quantized quantum spin Hall behavior. Moreover, at $\nu_{\rm tot}=-1$ we observe a well-defined resistance minimum when the middle and bottom layers are simultaneously half filled, a configuration naturally interpreted as an excitonic phase in the quantum Hall regime stabilized by interlayer coherence. Our results establish large-angle TTG as a platform where layer-resolved quantum Hall physics coexists with correlated phenomena that are unique to the trilayer geometry.

\begin{figure*}[!tbh]
\begin{center}
\includegraphics[width=\textwidth]{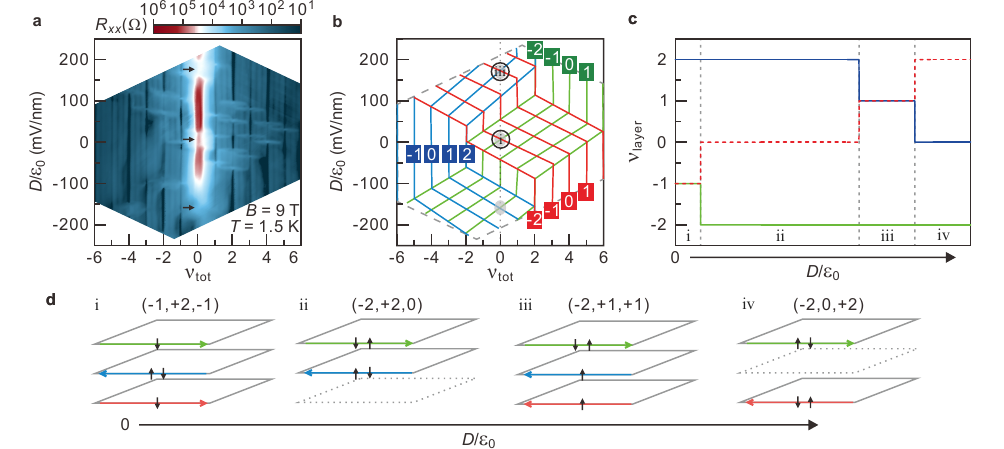}
\caption{{\bf Evolution of interlayer helical edge states with displacement field.} 
{\bf (a)} Longitudinal resistance $R_{xx}$ as a function of $\nu_{\rm tot}$ and $D/\varepsilon_{0}$ at $B = 9$~T. Three regions of reduced resistance appear along $\nu_{\rm tot}=0$ (arrows). {\bf (b)} Schematic representation of (a), showing the layer-resolved lowest Landau levels with the same color code in Fig.~\ref{Fig1}d, and the vertical splittings reflect broken spin and valley degeneracies. Numbers in the colored boxes mark the layer filling factors. The gray-shaded areas are the low-resistance regions at $\nu_{\rm tot}=0$. {\bf (c)} Hartree–Fock calculations of layer fillings at $\nu_{\rm tot}=0$ as a function of displacement field. Four stable configurations, labeled i–iv, appear as the displacement field is varied. {\bf (d)} Edge channel configurations corresponding to the four regions in (c). Black vertical arrows denote spin polarization and horizontal arrows indicate chirality.}

\label{Fig2}
\end{center}
\end{figure*}
The schematic of dual-gated large-angle TTG with alternating stacking device D1, shown in Fig.~\ref{Fig1}a, has a twist angle of about $5^\circ$. Two additional devices, D2 with alternating stacking and D3 with helical stacking, were also measured. Both devices exhibit overall magnetotransport features consistent with those of D1, and the corresponding datasets are provided in the Supporting Information. Figure~\ref{Fig1}b presents a color map of the longitudinal resistance $R_{xx}$ as a function of the displacement field ($D/\epsilon_0$) and the total filling factor ($\nu_\text{tot}$) at $B = 4$~T. Several phase transitions are visible, and within the range $-60 < \nu_\text{tot} < 60$ their positions are not symmetric about $\nu_\text{tot}=0$. This electron–hole asymmetry resembles that reported in dual-gated ABA trilayer graphene \cite{Taychatanapat2011, Henriksen2012PRX, Lee2013, Stepanov2016, Campos2016, Datta2017, Che2020, Chen2024}. In addition, diamond-shaped boundaries extend along the black dashed line in Fig.~\ref{Fig1}b, with further diamond features appearing symmetrically with respect to this line. To highlight these features, Fig.~\ref{Fig1}d provides a summary of the observed transitions. In this plot, the thick red, blue, and green curves denote the lowest Landau levels of the top, middle, and bottom layers, respectively, while the thinner traces correspond to higher-index Landau levels. Numbers in black boxes indicate the Landau level index $N$, and the dotted lines in Fig.~\ref{Fig1}d can be assigned to crossings between the top and bottom layer Landau levels with the same index.

In Fig.~\ref{Fig1}d, the vertical feature is attributed to the middle graphene layer, similar to earlier observations in chiral twisted trilayer graphene \cite{Davydov2025}. The screening effect of the outer layers suppresses the gate response of the middle layer, so its filling is less sensitive to the displacement field even as the outer layers change. This distinction causes the vertical features in Fig.~\ref{Fig1}b to be dominated by the middle layer, while the diagonal traces originate from the outer layers. Unlike the ideal ABA trilayer, where monolayer-like and bilayer-like subbands coexist, the large-angle TTG with alternating stacking does not show a bilayer-derived parabolic band. Instead, two Dirac-like dispersions originating from the top and bottom layers coexist with the Dirac cone of the middle sheet, so that the spectrum is effectively composed of three Dirac-like bands, as schematically shown in Fig.~\ref{Fig1}a. (See Fig.~S5(a) in the Supporting Information.) This difference is likely caused by residual misalignment between the outer graphene layers, which may arise from slight twist-angle deviations, interlayer sliding during assembly, or both.

To interpret the experimental results, we performed tight-binding calculations using the potential profile shown in the inset of Fig.~\ref{Fig1}a, based on the Slonczewski–Weiss–McClure parameterization commonly used to describe the monolayer-like Dirac bands in multilayer graphene. In this model the displacement field $D$ induces a potential difference $\Delta$ between the outer layers, while an additional parameter $\delta$ accounts for the residual charge imbalance between the outer and middle sheets. In the large-angle regime, the low-energy structure is well captured by three effectively decoupled Dirac bands subject to layer-dependent electrostatic potentials. The $\Delta$–$\delta$ parametrization implements this general description and does not rely on the microscopic twist orientation. For $D=0$, only $\delta$ is present, giving $V_{\rm top}=V_{\rm bot}=+\delta$ and $V_{\rm mid}=-2\delta$. With finite $D$, the layer potentials become
\[
V_{\rm top} = +\tfrac{\Delta}{2} + \delta,\quad 
V_{\rm mid} = -2\delta,\quad 
V_{\rm bot} = -\tfrac{\Delta}{2} + \delta.
\]

A small imbalance of order a few meV is typically required even in ABA trilayers, so we set $\delta=8$~meV \cite{Serbyn2013, Campos2016, Zibrov2018, Che2020}. With this choice the calculated density of states map in Fig.~\ref{Fig1}c reproduces the main features of the experiment. Circles highlight crossings of the lowest Landau levels, whose positions coincide with the transition points observed in the resistance map of Fig.~\ref{Fig1}b and summarized in Fig.~\ref{Fig1}d. The crossing of the first excited Landau levels predicted by the calculation also matches the corresponding transition in the data. This agreement shows that the potential model captures the essential electronic structure of the alternating-stacked, large-angle twisted trilayer graphene. Further results at higher magnetic fields are presented in Supplementary Note~2 and Supplementary Fig.~S5 in the Supporting Information.

At higher magnetic fields the effects of particle–hole asymmetry and charge imbalance become more pronounced. Figure~\ref{Fig2}a shows the longitudinal resistance in the lowest Landau level at $B=9$~T and $T=1.5$~K. Some fractional quantum Hall states are visible in the lowest Landau levels of the individual layers, and crossings between symmetry-broken states of different layers appear more clearly than at lower fields. The most striking feature occurs at $\nu_{\rm tot}=0$, where the resistance remains largely high across most displacement field, but three distinct regions of reduced resistance emerge, highlighted by arrows in Fig.~\ref{Fig2}a. (See Fig.~S2 in Supporting Information)

The schematic in Fig.~\ref{Fig2}b extends the picture of Fig.~\ref{Fig1}d, making explicit the symmetry-broken states that emerge under displacement-field tuning. Green, blue, and red lines denote the Landau levels of the top, middle, and bottom layers, respectively, and their vertical splittings reflect the lifting of spin and valley degeneracies. Numbers in the colored boxes indicate the layer filling factors, and the three shaded regions correspond to $(\nu_{\rm top},\nu_{\rm mid},\nu_{\rm bot})=(-2,1,1)$, $(-1,2,-1)$, and $(1,1,-2)$ for positive, near-zero, and negative displacement fields. These configurations match the positions of the resistance minima at $\nu_{\rm tot}=0$ in Fig.~\ref{Fig2}a, showing that the low-resistance states originate from specific combinations of symmetry-broken Landau levels across the three layers.

These symmetry-breaking patterns are further supported by our mean-field Hartree–Fock calculations at $\nu_{\rm tot}=0$, which track the layer-resolved filling factors as a function of displacement field, as shown in Fig.~\ref{Fig2}c. In this calculation, we model each layer as continuum Dirac fermions subject to a uniform magnetic field, with the asymmetry $\delta$ determined from our tight-binding analysis (Supplementary Note~3). The three colored traces give $\nu_{\rm top}$, $\nu_{\rm mid}$, and $\nu_{\rm bot}$, whose sum is constrained to zero at all fields. Four different configurations, labeled i–iv, appear as the displacement field is swept. In direct comparison with the data in Fig.~\ref{Fig2}a, states ii and iv coincide with high-resistance sectors, while states i and iii align with the low-resistance regions at $\nu_{\rm tot}=0$.

\begin{figure}
\begin{center}
\includegraphics{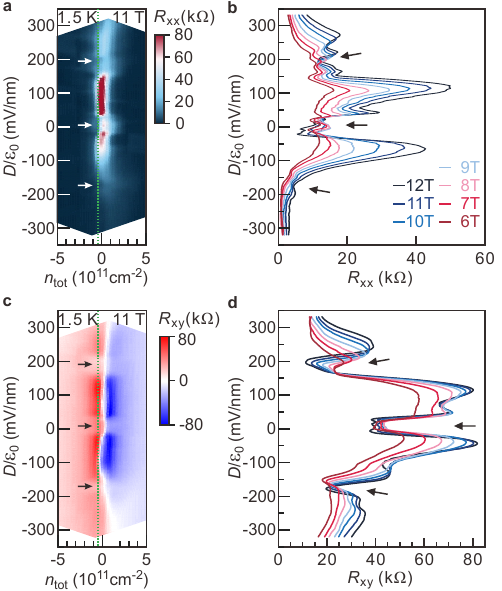}
\caption{{\bf Magnetic field dependence of helical edge states.} {\bf (a)} Contour map of longitudinal resistance $R_{xx}$ as a function of carrier density and displacement field at $B=11$~T. {\bf (b)} Line traces of $R_{xx}$ as a function of displacement field at $n_{\rm tot}=5\times10^{10}\,\mathrm{cm}^{-2}$, taken along the green dotted line in (a), for magnetic fields between 12~T and 6~T. 
{\bf (c)} Contour map of Hall resistance $R_{xy}$ measured under the same conditions as in (a). {\bf (d)} Line traces of $R_{xy}$ corresponding to the green dotted line in (c), at $n_{\rm tot}=5\times10^{10}\,\mathrm{cm}^{-2}$ for fields between 12~T and 6~T.}
\label{Fig3}
\end{center}
\end{figure}

The underlying edge-channel configurations of these correlated insulators are illustrated in Fig.~\ref{Fig2}d. Left- (right-) pointing arrows indicate electron-like (hole-like) chirality for each layer-resolved edge mode. In the high-resistance cases (ii and iv), the relevant fillings include $\pm 2$ on adjacent layers, which generate counter-propagating channels of the same spin; interlayer scattering is then efficient and $R_{xx}$ is enhanced~\cite{SanchezYamagishi2017,Hoke2024}. In contrast, in configurations i and iii the numbers of right- and left-moving channels for a fixed spin do not match. For example, in case i the left-moving spin-$\uparrow$ mode has no same-spin partner for scattering, while the spin-$\downarrow$ channels equilibrate among themselves. Because spin-$z$ is approximately conserved in graphene, inter-spin scattering is strongly suppressed, leaving one spin-$\uparrow$ branch and one spin-$\downarrow$ branch that do not efficiently backscatter. This spin-selective equilibration reduces $R_{xx}$ and yields a helical-like edge configuration in the quantum Hall regime. The Hall response in Fig.~\ref{Fig3} shows suppressed $R_{xy}$ at these fillings, consistent with this spin-resolved helical picture under broken time-reversal symmetry.

Such spin-resolved interlayer helical edge configurations are a characteristic feature of large-angle TTG, realized for example by the filling sequence $(\nu_{\rm top},\nu_{\rm mid},\nu_{\rm bot})=(\pm1,\pm1,\mp2)$. Similar $\nu_{\rm tot}=0$ low-resistance states are observed in both device~D2 and device~D3, which have alternating-like and helical stacking, respectively. When $\nu_{\rm tot}=0$, the numbers of electron- and hole-like channels compensate, so counter-propagating modes with opposite spin can form a helical-like edge configuration owing to the approximate conservation of spin-$z$ in graphene under magnetic fields. For $\nu_{\rm tot}\neq0$, this compensation is lost, and an additional co-propagating channel provides an equilibration pathway that renders such configurations fragile. This explains why analogous states have not been observed in bilayers: in twisted bilayer graphene the $(-1,+1)$ configuration at $\nu_{\rm tot}=0$ can host a related state~\cite{SanchezYamagishi2017}, but the $(-2,+1)$ configuration lacks a partner channel and reduces to a conventional quantum Hall state. In large-angle TTG the additional $\nu=+1$ layer supplies this partner, yielding spin-resolved helical and non-helical edge modes that remain detectable over a finite range of displacement field.

To further confirm this interpretation, we carried out non-local transport measurements, which probe the helical-like character of the edge channels. The details of these measurements, including probe configurations and field dependence, are provided in Supplementary Note~4, Fig.~S6 and Table~S2.

The magnetic field dependence of the helical-like edge configurations is summarized in Fig.~\ref{Fig3}. At $B=11$~T, the longitudinal resistance map in Fig.~\ref{Fig3}a shows clear reductions in $R_{xx}$ along $\nu_{\rm tot}=0$, consistent with the features identified in Fig.~\ref{Fig2}. These reductions are most prominent in the filling combinations $(-2,1,1)$, $(-1,2,-1)$, and $(1,1,-2)$, indicated by the white arrows. Additional suppression appears for $(\pm2,0,\mp2)$ compared with $(0,\pm2,\mp2)$, which is consistent with the larger spacing between the relevant edge channels reducing interlayer backscattering. 

Line cuts taken at $n_{\rm tot}=5\times10^{10}\,\mathrm{cm}^{-2}$, shown in Fig.~\ref{Fig3}b, demonstrate that the dips in $R_{xx}$ persist from $B=12$~T down to 6~T, but the dip becomes less pronounced relative to the surrounding background, indicating reduced contrast. We note, however, that the same displacement-field windows remain clearly identifiable in the non-local response, where the contrast is more robust, as shown in Fig.~S6 of the Supporting Information. This trend reflects the role of field-induced spin and valley splitting, which stabilizes the helical-like configurations and enters the Hartree–Fock analysis of Fig.~\ref{Fig2}c.

The Hall resistance exhibits the same evolution. In Fig.~\ref{Fig3}c, the contour map at 11~T shows reduced $R_{xy}$ in the same displacement-field windows where $R_{xx}$ is suppressed. The corresponding line cuts in Fig.~\ref{Fig3}d reveal that the dip becomes less distinct relative to the surrounding background at lower fields, mirroring the trend in the longitudinal response. As the magnetic field decreases, the spin and valley splitting of the $N=0$ Landau level becomes weaker, interlayer scattering is restored, and the low-resistance states gradually disappear.

\begin{figure}
\begin{center}
\includegraphics{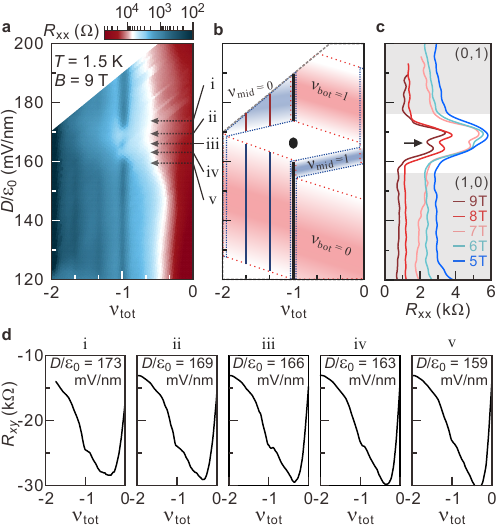}
\caption{{\bf Interlayer coherence quantum Hall states at $\nu_{\rm tot}=-1$.} {\bf (a)} Longitudinal resistance $R_{xx}$ plotted as a function of $\nu_{\rm tot}$ and $D/\varepsilon_{0}$. {\bf (b)} Schematic map of (a). Blue and red shading denotes filling of the middle and bottom layers, respectively, while the top layer remains fixed at $\nu_{\rm top}=-2$. The black solid line at $\nu_{\rm tot}=-1$ corresponds to integer quantum Hall states with $(\nu_{\rm mid},\nu_{\rm bot})=(0,1)$ and $(1,0)$. Dark red (blue) lines in the interval $-2<\nu_{\rm tot}<-1$ indicate fractional quantum Hall states of the bottom (middle) layer. The black dot marks a low-resistance region. {\bf (c)} Evolution of $R_{xx}$ with displacement field at $\nu_{\rm tot}=-1$ for several magnetic fields. The arrow highlights resistivity minima. Curves are offset vertically and horizontally for clarity. {\bf (d)} Line traces of $R_{xy}$ at various fixed $D/\varepsilon_{0}$ values indicated by (i-v) in panel (a).}
\label{Fig4}
\end{center}
\end{figure}

Another intriguing state of large-angle TTG emerges at $\nu_{\rm tot}=-1$, where an interlayer excitonic state develops. Figure~\ref{Fig4}a presents a magnified window of the $(\nu_{\rm tot},D/\varepsilon_0)$ plane over the range $-2<\nu_{\rm tot}<0$. In this regime the top layer is locked at $\nu_{\rm top}=-2$, leaving the middle and bottom sheets to host the relevant lowest Landau level physics. Fractional quantum Hall features at $\nu=-5/3$ and $-4/3$ arise from one of these two active layers, as summarized in Fig.~\ref{Fig4}b.

At $\nu_{\rm tot}=-1$ and $D/\varepsilon_0\approx 166$~mV/nm, a narrow dip develops in $R_{xx}$. The schematic in Fig.~\ref{Fig4}b identifies this point as the balanced configuration $\nu_{\rm mid}=\nu_{\rm bot}=1/2$, corresponding to half filling of the lowest Landau level in both adjacent layers. This is the standard condition for interlayer coherent exciton formation in graphene bilayers, realized here without an insulating barrier between the two active sheets~\cite{Kim2021,Kim2023,Kim2025,Kim2023AdvSci,Li2024}. At this excitonic condition, $R_{xy}$ remains close to the expected $\nu_{\rm tot}=-1$ quantum Hall plateau value (Fig.~\ref{Fig4}d-iii), and as the displacement field is tuned away, the $\nu_{\rm tot}=-1$ plateau becomes less well defined (Fig.~\ref{Fig4}d-ii and iv). Although interlayer coherent phases can in principle emerge whenever two adjacent layers are simultaneously half filled, in our device a clean and well-isolated half-filling condition is realized only at $\nu_{\rm tot}=-1$. These observations highlight how a hallmark of bilayer quantum Hall physics,interlayer exciton condensation at half filling, acquires a new realization in twisted trilayers, in which one layer remains inert while the remaining two condense into an excitonic phase. The observed minimum in $R_{xx}$ can therefore be understood as the combined response of a conventional edge channel in the top layer and a coherent electron–hole condensate in the middle–bottom subsystem, showing how conventional quantum Hall order and interlayer correlated phases intertwine in a trilayer platform enabled by large-angle decoupling.

The field dependence of this feature is shown in Fig.~\ref{Fig4}c. Line traces of $R_{xx}$ at $\nu_{\rm tot}=-1$ reveal that the dip (black arrows) becomes progressively weaker as the magnetic field is lowered from 9~T to 5~T. This trend is consistent with an interaction-driven coherent state whose stability depends on the exchange-enhanced gap at half filling: as the field decreases, the gap diminishes and the excitonic signature fades~\cite{Kim2023,Kim2025,Kim2023AdvSci,Li2024}.

In conclusion, we have investigated the quantum Hall states of large-angle TTG and identified correlated interlayer quantum Hall phases that are not accessible in conventional bilayer and trilayer systems. The middle layer responds primarily to carrier density while the outer layers are strongly influenced by both density and displacement field, producing a landscape of filling-factor combinations characteristic of this trilayer geometry. At $\nu_{\rm tot}=0$, a reduction in resistance appears. Comparison with Hartree–Fock calculations indicates that these low-resistance states arise from spin-resolved helical-like edge configurations in the quantum Hall regime, rather than from a fully quantized quantum spin Hall state. In contrast to the bilayer case, all three graphene layers participate, enabling configurations such as $(\pm2,0,\mp2)$ and $(0,\pm2,\mp2)$, which illustrate how interlayer separation influences the extent of backscattering. This shows that large-angle TTG offers a controllable setting for examining spin-polarized edge transport in a multilayer environment.

A further characteristic of this system is the emergence of an interlayer excitonic phase at $\nu_{\rm tot}=-1$. In this regime the top sheet forms a conventional integer quantum Hall state at $\nu_{\rm top}=-2$, while the middle and bottom sheets are simultaneously half filled, creating conditions for excitonic coherence across adjacent layers. The resulting suppression of $R_{xx}$ resembles excitonic behavior reported previously in twisted bilayer and double bilayer graphene, but here it coexists with a fully filled outer layer, providing a trilayer realization of this phenomenon. The magnetic-field dependence of the feature is consistent with an interaction-driven origin. In summary, large-angle TTG combines conventional quantum Hall features with interlayer correlated phases, illustrating how trilayer systems with tunable layer asymmetry can host spin-resolved edge transport and excitonic order.

\section{Supporting Information}
Methods, Additional Experimental Data, Band structure calculations, Hartree-Fock calculations, Non-local measurement

\section{Acknowledgement} 
The work from DGIST was supported by the National Research Foundation of Korea (NRF) (Grant No. RS-2025-00557717, RS-2023-00269616)  and the Nano and Material Technology Development Program through the National Research Foundation of Korea (NRF) funded by Ministry of Science and ICT (No. RS-2024-00444725). We also acknowledge the partner group program of the Max Planck Society. Part of this work was supported by Global Partnership Program of Leading Universities in Quantum Science and Technology (RS-2025-02317602). G. Y. C. is supported by Samsung Science and Technology Foundation under Project Number SSTF-BA2401-03, the NRF of Korea (Grants No. RS-2023-00208291, RS-2024-00410027, 2023M3K5A1094810, RS-2023-NR119931, RS-2024-00444725, RS-2023-00256050, IRS-2025-25453111, RS-2025-08542968) funded by the Korean Government (MSIT), the Air Force Office of Scientific Research under Award No. FA23862514026, and Institute of Basic Science under project code IBS-R014-D1. K.W. and T.T. acknowledge support from the JSPS KAKENHI (Grant Numbers 21H05233 and 23H02052) , the CREST (JPMJCR24A5), JST and World Premier International Research Center Initiative (WPI), MEXT, Japan. This work at UOS was supported by the National Research Foundation of Korea (NRF) through Grant No. RS-2023-00249414 (N.L.) and Grant No. NRF-2020R1A5A1016518 (J.J.). Computational resources were provided by KISTI under Grant No. KSC-2022-CRE-0514 and by the Urban Big Data and AI Institute (UBAI) at UOS.

\section{References} 



\section{Author contributions} 
D.K., Y.K., and G.Y.C. conceived the project. D.K. fabricated the devices and carried out low-temperature measurements with Y.K. Hartree–Fock calculations were performed by G.L., S.J., and G.Y.C. Tight-binding calculations by N.L. and J.J. T.T. and K.W. synthesized the h-BN crystals. All authors contributed to the writing of the manuscript. D.K. and G.L. contributed equally to this work.

\section{ Competing interests} 
The authors declare no competing interest.


\begin{thebibliography}{31}%
\makeatletter
\providecommand \@ifxundefined [1]{%
 \@ifx{#1\undefined}
}%
\providecommand \@ifnum [1]{%
 \ifnum #1\expandafter \@firstoftwo
 \else \expandafter \@secondoftwo
 \fi
}%
\providecommand \@ifx [1]{%
 \ifx #1\expandafter \@firstoftwo
 \else \expandafter \@secondoftwo
 \fi
}%
\providecommand \natexlab [1]{#1}%
\providecommand \enquote  [1]{``#1''}%
\providecommand \bibnamefont  [1]{#1}%
\providecommand \bibfnamefont [1]{#1}%
\providecommand \citenamefont [1]{#1}%
\providecommand \href@noop [0]{\@secondoftwo}%
\providecommand \href [0]{\begingroup \@sanitize@url \@href}%
\providecommand \@href[1]{\@@startlink{#1}\@@href}%
\providecommand \@@href[1]{\endgroup#1\@@endlink}%
\providecommand \@sanitize@url [0]{\catcode `\\12\catcode `\$12\catcode `\&12\catcode `\#12\catcode `\^12\catcode `\_12\catcode `\%12\relax}%
\providecommand \@@startlink[1]{}%
\providecommand \@@endlink[0]{}%
\providecommand \url  [0]{\begingroup\@sanitize@url \@url }%
\providecommand \@url [1]{\endgroup\@href {#1}{\urlprefix }}%
\providecommand \urlprefix  [0]{URL}%
\providecommand \Eprint [0]{\href}%
\providecommand \doibase [0]{https://doi.org/}%
\providecommand \selectlanguage [0]{\@gobble}%
\providecommand \bibinfo  [0]{\@secondoftwo}%
\providecommand \bibfield  [0]{\@secondoftwo}%
\providecommand \translation [1]{[#1]}%
\providecommand \BibitemOpen [0]{}%
\providecommand \bibitemStop [0]{}%
\providecommand \bibitemNoStop [0]{.\EOS\space}%
\providecommand \EOS [0]{\spacefactor3000\relax}%
\providecommand \BibitemShut  [1]{\csname bibitem#1\endcsname}%
\let\auto@bib@innerbib\@empty

\bibitem [{\citenamefont {Koshino}\ and\ \citenamefont {McCann}(2011)}]{Koshino2011}%
  \BibitemOpen
  \bibfield  {author} {\bibinfo {author} {\bibfnamefont {M.}\ \bibnamefont {Koshino}},\ and\ 
  \bibinfo {author} {\bibfnamefont {E.}\ \bibnamefont {McCann}},\ }\bibfield  {title} 
  {\bibinfo {title} {Landau level spectra and the quantum Hall effect of multilayer graphene},\ }
  \href {https://doi.org/10.1103/PhysRevB.83.165443} {\bibfield  {journal} {\bibinfo  {journal} {Phys. Rev. B}\ }\textbf {\bibinfo {volume} {83}},\ \bibinfo {pages} {165443} (\bibinfo {year} {2011})}\BibitemShut {NoStop}%
\bibitem [{\citenamefont {Taychatanapat}\ \emph {et~al.}(2011)\citenamefont {Taychatanapat}, \citenamefont {Watanabe}, \citenamefont {Taniguchi},\ and\ \citenamefont {Jarillo-Herrero}}]{Taychatanapat2011}%
  \BibitemOpen
  \bibfield  {author} {\bibinfo {author} {\bibfnamefont {T.}\ \bibnamefont {Taychatanapat}}, 
  \bibinfo {author} {\bibfnamefont {K.}\ \bibnamefont {Watanabe}}, 
  \bibinfo {author} {\bibfnamefont {T.}\ \bibnamefont {Taniguchi}},\ and\ 
  \bibinfo {author} {\bibfnamefont {P.}\ \bibnamefont {Jarillo-Herrero}},\ }\bibfield  {title} 
  {\bibinfo {title} {Quantum Hall effect and Landau-level crossing of Dirac fermions in trilayer graphene},\ }
  \href {https://doi.org/10.1038/nphys2008} {\bibfield  {journal} {\bibinfo  {journal} {Nat. Phys.}\ }\textbf {\bibinfo {volume} {7}},\ \bibinfo {pages} {621--625} (\bibinfo {year} {2011})}\BibitemShut {NoStop}%
\bibitem [{\citenamefont {Henriksen}, \citenamefont {Nandi},\ and\ \citenamefont {Eisenstein}(2012)}]{Henriksen2012PRX}%
  \BibitemOpen
  \bibfield  {author} {\bibinfo {author} {\bibfnamefont {E.~A.}\ \bibnamefont {Henriksen}}, 
  \bibinfo {author} {\bibfnamefont {D.}\ \bibnamefont {Nandi}},\ and\ 
  \bibinfo {author} {\bibfnamefont {J.~P.}\ \bibnamefont {Eisenstein}},\ }\bibfield  {title} 
  {\bibinfo {title} {Quantum Hall Effect and Semimetallic Behavior of Dual-Gated {ABA}-Stacked Trilayer Graphene},\ }
  \href {https://doi.org/10.1103/PhysRevX.2.011004} {\bibfield  {journal} {\bibinfo  {journal} {Phys. Rev. X}\ }\textbf {\bibinfo {volume} {2}},\ \bibinfo {pages} {011004} (\bibinfo {year} {2012})}\BibitemShut {NoStop}%
\bibitem [{\citenamefont {Serbyn}\ and\ \citenamefont {Abanin}(2013)}]{Serbyn2013}%
  \BibitemOpen
  \bibfield  {author} {\bibinfo {author} {\bibfnamefont {M.}\ \bibnamefont {Serbyn}}\ and\ 
  \bibinfo {author} {\bibfnamefont {D.~A.}\ \bibnamefont {Abanin}},\ }\bibfield  {title} 
  {\bibinfo {title} {New Dirac points and multiple Landau level crossings in biased trilayer graphene},\ }
  \href {https://doi.org/10.1103/PhysRevB.87.115422} {\bibfield  {journal} {\bibinfo  {journal} {Phys. Rev. B}\ }\textbf {\bibinfo {volume} {87}},\ \bibinfo {pages} {115422} (\bibinfo {year} {2013})}\BibitemShut {NoStop}%
\bibitem [{\citenamefont {Lee}\ \emph {et~al.}(2013)\citenamefont {Lee}, \citenamefont {Velasco~Jr.}, \citenamefont {Tran}, \citenamefont {Zhang}, \citenamefont {Bao}, \citenamefont {Jing}, \citenamefont {Myhro}, \citenamefont {Smirnov},\ and\ \citenamefont {Lau}}]{Lee2013}%
  \BibitemOpen
  \bibfield  {author} {\bibinfo {author} {\bibfnamefont {Y.}\ \bibnamefont {Lee}},
  \bibinfo {author} {\bibfnamefont {J.}\ \bibnamefont {Velasco~Jr.}},
  \bibinfo {author} {\bibfnamefont {D.}\ \bibnamefont {Tran}},
  \bibinfo {author} {\bibfnamefont {F.}\ \bibnamefont {Zhang}},
  \bibinfo {author} {\bibfnamefont {W.}\ \bibnamefont {Bao}},
  \bibinfo {author} {\bibfnamefont {L.}\ \bibnamefont {Jing}},
  \bibinfo {author} {\bibfnamefont {K.}\ \bibnamefont {Myhro}},
  \bibinfo {author} {\bibfnamefont {D.}\ \bibnamefont {Smirnov}},\ and\
  \bibinfo {author} {\bibfnamefont {C.~N.}\ \bibnamefont {Lau}},\ }\bibfield  {title} 
  {\bibinfo {title} {Broken symmetry quantum Hall states in dual-gated {ABA} trilayer graphene},\ }
  \href {https://doi.org/10.1021/nl4000757} {\bibfield  {journal} {\bibinfo  {journal} {Nano Lett.}\ }\textbf {\bibinfo {volume} {13}},\ \bibinfo {pages} {1627--1631} (\bibinfo {year} {2013})}\BibitemShut {NoStop}%
\bibitem [{\citenamefont {Campos}\ \emph {et~al.}(2016)\citenamefont {Campos}, \citenamefont {Taychatanapat}, \citenamefont {Serbyn}, \citenamefont {Surakitbovorn}, \citenamefont {Watanabe}, \citenamefont {Taniguchi}, \citenamefont {Abanin},\ and\ \citenamefont {Jarillo-Herrero}}]{Campos2016}%
  \BibitemOpen
  \bibfield  {author} {\bibinfo {author} {\bibfnamefont {L.~C.}\ \bibnamefont {Campos}},
  \bibinfo {author} {\bibfnamefont {T.}\ \bibnamefont {Taychatanapat}},
  \bibinfo {author} {\bibfnamefont {M.}\ \bibnamefont {Serbyn}},
  \bibinfo {author} {\bibfnamefont {K.}\ \bibnamefont {Surakitbovorn}},
  \bibinfo {author} {\bibfnamefont {K.}\ \bibnamefont {Watanabe}},
  \bibinfo {author} {\bibfnamefont {T.}\ \bibnamefont {Taniguchi}},
  \bibinfo {author} {\bibfnamefont {D.~A.}\ \bibnamefont {Abanin}},\ and\
  \bibinfo {author} {\bibfnamefont {P.}\ \bibnamefont {Jarillo-Herrero}},\ }\bibfield  {title} 
  {\bibinfo {title} {Landau level splittings, phase transitions, and nonuniform charge distribution in trilayer graphene},\ }
  \href {https://doi.org/10.1103/PhysRevLett.117.066601} {\bibfield  {journal} {\bibinfo  {journal} {Phys. Rev. Lett.}\ }\textbf {\bibinfo {volume} {117}},\ \bibinfo {pages} {066601} (\bibinfo {year} {2016})}\BibitemShut {NoStop}%
\bibitem [{\citenamefont {Stepanov}\ \emph {et~al.}(2016)\citenamefont {Stepanov}, \citenamefont {Barlas}, \citenamefont {Espiritu}, \citenamefont {Che}, \citenamefont {Watanabe}, \citenamefont {Taniguchi}, \citenamefont {Smirnov},\ and\ \citenamefont {Lau}}]{Stepanov2016}%
  \BibitemOpen
  \bibfield  {author} {\bibinfo {author} {\bibfnamefont {P.}\ \bibnamefont {Stepanov}},
  \bibinfo {author} {\bibfnamefont {Y.}\ \bibnamefont {Barlas}},
  \bibinfo {author} {\bibfnamefont {T.}\ \bibnamefont {Espiritu}},
  \bibinfo {author} {\bibfnamefont {S.}\ \bibnamefont {Che}},
  \bibinfo {author} {\bibfnamefont {K.}\ \bibnamefont {Watanabe}},
  \bibinfo {author} {\bibfnamefont {T.}\ \bibnamefont {Taniguchi}},
  \bibinfo {author} {\bibfnamefont {D.}\ \bibnamefont {Smirnov}},\ and\
  \bibinfo {author} {\bibfnamefont {C.~N.}\ \bibnamefont {Lau}},\ }\bibfield  {title} 
  {\bibinfo {title} {Tunable symmetries of integer and fractional quantum Hall phases in heterostructures with multiple Dirac bands},\ }
  \href {https://doi.org/10.1103/PhysRevLett.117.076807} {\bibfield  {journal} {\bibinfo  {journal} {Phys. Rev. Lett.}\ }\textbf {\bibinfo {volume} {117}},\ \bibinfo {pages} {076807} (\bibinfo {year} {2016})}\BibitemShut {NoStop}%
\bibitem [{\citenamefont {Datta}\ \emph {et~al.}(2017)\citenamefont {Datta}, \citenamefont {Dey}, \citenamefont {Samanta}, \citenamefont {Agarwal}, \citenamefont {Borah}, \citenamefont {Watanabe}, \citenamefont {Taniguchi}, \citenamefont {Sensarma},\ and\ \citenamefont {Deshmukh}}]{Datta2017}%
  \BibitemOpen
  \bibfield  {author} {\bibinfo {author} {\bibfnamefont {B.}\ \bibnamefont {Datta}},
  \bibinfo {author} {\bibfnamefont {S.}\ \bibnamefont {Dey}},
  \bibinfo {author} {\bibfnamefont {A.}\ \bibnamefont {Samanta}},
  \bibinfo {author} {\bibfnamefont {H.}\ \bibnamefont {Agarwal}},
  \bibinfo {author} {\bibfnamefont {A.}\ \bibnamefont {Borah}},
  \bibinfo {author} {\bibfnamefont {K.}\ \bibnamefont {Watanabe}},
  \bibinfo {author} {\bibfnamefont {T.}\ \bibnamefont {Taniguchi}},
  \bibinfo {author} {\bibfnamefont {R.}\ \bibnamefont {Sensarma}},\ and\
  \bibinfo {author} {\bibfnamefont {M.~M.}\ \bibnamefont {Deshmukh}},\ }\bibfield  {title} 
  {\bibinfo {title} {Strong electronic interaction and multiple quantum Hall ferromagnetic phases in trilayer graphene},\ }
  \href {https://doi.org/10.1038/ncomms14518} {\bibfield  {journal} {\bibinfo  {journal} {Nat. Commun.}\ }\textbf {\bibinfo {volume} {8}},\ \bibinfo {pages} {14518} (\bibinfo {year} {2017})}\BibitemShut {NoStop}%
\bibitem [{\citenamefont {Zibrov}\ \emph {et~al.}(2018)\citenamefont {Zibrov}, \citenamefont {Rao}, \citenamefont {Kometter}, \citenamefont {Spanton}, \citenamefont {Li}, \citenamefont {Dean}, \citenamefont {Taniguchi}, \citenamefont {Watanabe}, \citenamefont {Serbyn},\ and\ \citenamefont {Young}}]{Zibrov2018}%
  \BibitemOpen
  \bibfield  {author} {\bibinfo {author} {\bibfnamefont {A.~A.}\ \bibnamefont {Zibrov}},
  \bibinfo {author} {\bibfnamefont {P.}\ \bibnamefont {Rao}},
  \bibinfo {author} {\bibfnamefont {C.}\ \bibnamefont {Kometter}},
  \bibinfo {author} {\bibfnamefont {E.~M.}\ \bibnamefont {Spanton}},
  \bibinfo {author} {\bibfnamefont {J.~I.~A.}\ \bibnamefont {Li}},
  \bibinfo {author} {\bibfnamefont {C.~R.}\ \bibnamefont {Dean}},
  \bibinfo {author} {\bibfnamefont {T.}\ \bibnamefont {Taniguchi}},
  \bibinfo {author} {\bibfnamefont {K.}\ \bibnamefont {Watanabe}},
  \bibinfo {author} {\bibfnamefont {M.}\ \bibnamefont {Serbyn}},\ and\
  \bibinfo {author} {\bibfnamefont {A.~F.}\ \bibnamefont {Young}},\ }\bibfield  {title} 
  {\bibinfo {title} {Emergent Dirac gullies and gully-symmetry-breaking quantum Hall states in {ABA} trilayer graphene},\ }
  \href {https://doi.org/10.1103/PhysRevLett.121.167601} {\bibfield  {journal} {\bibinfo  {journal} {Phys. Rev. Lett.}\ }\textbf {\bibinfo {volume} {121}},\ \bibinfo {pages} {167601} (\bibinfo {year} {2018})}\BibitemShut {NoStop}%
\bibitem [{\citenamefont {Che}\ \emph {et~al.}(2020)\citenamefont {Che}, \citenamefont {Stepanov}, \citenamefont {Ge}, \citenamefont {Zhu}, \citenamefont {Wang}, \citenamefont {Lee}, \citenamefont {Myhro}, \citenamefont {Shi}, \citenamefont {Chen}, \citenamefont {Pi}, \citenamefont {Pan}, \citenamefont {Cheng}, \citenamefont {Taniguchi}, \citenamefont {Watanabe}, \citenamefont {Barlas}, \citenamefont {Lake}, \citenamefont {Bockrath}, \citenamefont {Hwang},\ and\ \citenamefont {Lau}}]{Che2020}%
  \BibitemOpen
  \bibfield  {author} {\bibinfo {author} {\bibfnamefont {S.}\ \bibnamefont {Che}},
  \bibinfo {author} {\bibfnamefont {P.}\ \bibnamefont {Stepanov}},
  \bibinfo {author} {\bibfnamefont {S.}\ \bibnamefont {Ge}},
  \bibinfo {author} {\bibfnamefont {M.}\ \bibnamefont {Zhu}},
  \bibinfo {author} {\bibfnamefont {D.}\ \bibnamefont {Wang}},
  \bibinfo {author} {\bibfnamefont {Y.}\ \bibnamefont {Lee}},
  \bibinfo {author} {\bibfnamefont {K.}\ \bibnamefont {Myhro}},
  \bibinfo {author} {\bibfnamefont {Y.}\ \bibnamefont {Shi}},
  \bibinfo {author} {\bibfnamefont {R.}\ \bibnamefont {Chen}},
  \bibinfo {author} {\bibfnamefont {Z.}\ \bibnamefont {Pi}},
  \bibinfo {author} {\bibfnamefont {C.}\ \bibnamefont {Pan}},
  \bibinfo {author} {\bibfnamefont {B.}\ \bibnamefont {Cheng}},
  \bibinfo {author} {\bibfnamefont {T.}\ \bibnamefont {Taniguchi}},
  \bibinfo {author} {\bibfnamefont {K.}\ \bibnamefont {Watanabe}},
  \bibinfo {author} {\bibfnamefont {Y.}\ \bibnamefont {Barlas}},
  \bibinfo {author} {\bibfnamefont {R.~K.}\ \bibnamefont {Lake}},
  \bibinfo {author} {\bibfnamefont {M.}\ \bibnamefont {Bockrath}},
  \bibinfo {author} {\bibfnamefont {J.}\ \bibnamefont {Hwang}},\ and\
  \bibinfo {author} {\bibfnamefont {C.~N.}\ \bibnamefont {Lau}},\ }\bibfield  {title} 
  {\bibinfo {title} {Substrate-dependent band structures in trilayer graphene/h-BN heterostructures},\ }
  \href {https://doi.org/10.1103/PhysRevLett.125.246401} {\bibfield  {journal} {\bibinfo  {journal} {Phys. Rev. Lett.}\ }\textbf {\bibinfo {volume} {125}},\ \bibinfo {pages} {246401} (\bibinfo {year} {2020})}\BibitemShut {NoStop}%
\bibitem [{\citenamefont {Chen}\ \emph {et~al.}(2024)\citenamefont {Chen}, \citenamefont {Huang}, \citenamefont {Li}, \citenamefont {Tong}, \citenamefont {Kuang}, \citenamefont {Xi}, \citenamefont {Watanabe}, \citenamefont {Taniguchi}, \citenamefont {Liu}, \citenamefont {Zhu}, \citenamefont {Lu}, \citenamefont {Zhang}, \citenamefont {Wu},\ and\ \citenamefont {Wang}}]{Chen2024}%
  \BibitemOpen
  \bibfield  {author} {\bibinfo {author} {\bibfnamefont {Y.}\ \bibnamefont {Chen}},
  \bibinfo {author} {\bibfnamefont {Y.}\ \bibnamefont {Huang}},
  \bibinfo {author} {\bibfnamefont {Q.}\ \bibnamefont {Li}},
  \bibinfo {author} {\bibfnamefont {B.}\ \bibnamefont {Tong}},
  \bibinfo {author} {\bibfnamefont {G.}\ \bibnamefont {Kuang}},
  \bibinfo {author} {\bibfnamefont {C.}\ \bibnamefont {Xi}},
  \bibinfo {author} {\bibfnamefont {K.}\ \bibnamefont {Watanabe}},
  \bibinfo {author} {\bibfnamefont {T.}\ \bibnamefont {Taniguchi}},
  \bibinfo {author} {\bibfnamefont {G.}\ \bibnamefont {Liu}},
  \bibinfo {author} {\bibfnamefont {Z.}\ \bibnamefont {Zhu}},
  \bibinfo {author} {\bibfnamefont {L.}\ \bibnamefont {Lu}},
  \bibinfo {author} {\bibfnamefont {F.-C.}\ \bibnamefont {Zhang}},
  \bibinfo {author} {\bibfnamefont {Y.-H.}\ \bibnamefont {Wu}},\ and\ 
  \bibinfo {author} {\bibfnamefont {L.}\ \bibnamefont {Wang}},\ }\bibfield  {title} 
  {\bibinfo {title} {Tunable even- and odd-denominator fractional quantum Hall states in trilayer graphene},\ }
  \href {https://doi.org/10.1038/s41467-024-50589-2} {\bibfield  {journal} {\bibinfo  {journal} {Nat. Commun.}\ }\textbf {\bibinfo {volume} {15}},\ \bibinfo {pages} {6236} (\bibinfo {year} {2024})}\BibitemShut {NoStop}%
\bibitem [{\citenamefont {Koshino}\ and\ \citenamefont {McCann}(2009)}]{Koshino2009}%
  \BibitemOpen
  \bibfield  {author} {\bibinfo {author} {\bibfnamefont {M.}\ \bibnamefont {Koshino}},\ and\ 
  \bibinfo {author} {\bibfnamefont {E.}\ \bibnamefont {McCann}},\ }\bibfield  {title} 
  {\bibinfo {title} {Trigonal warping and Berry's phase $N\pi$ in {ABC}-stacked multilayer graphene},\ }
  \href {https://doi.org/10.1103/PhysRevB.80.165409} {\bibfield  {journal} {\bibinfo  {journal} {Phys. Rev. B}\ }\textbf {\bibinfo {volume} {80}},\ \bibinfo {pages} {165409} (\bibinfo {year} {2009})}\BibitemShut {NoStop}%
\bibitem [{\citenamefont {Zhang}\ \emph {et~al.}(2010)\citenamefont {Zhang}, \citenamefont {Sahu}, \citenamefont {Min},\ and\ \citenamefont {MacDonald}}]{Zhang2010}%
  \BibitemOpen
  \bibfield  {author} {\bibinfo {author} {\bibfnamefont {F.}\ \bibnamefont {Zhang}}, 
  \bibinfo {author} {\bibfnamefont {B.}\ \bibnamefont {Sahu}}, 
  \bibinfo {author} {\bibfnamefont {H.}\ \bibnamefont {Min}},\ and\ 
  \bibinfo {author} {\bibfnamefont {A.~H.}\ \bibnamefont {MacDonald}},\ }\bibfield  {title} 
  {\bibinfo {title} {Band structure of {ABC}-stacked graphene trilayers},\ }
  \href {https://doi.org/10.1103/PhysRevB.82.035409} {\bibfield  {journal} {\bibinfo  {journal} {Phys. Rev. B}\ }\textbf {\bibinfo {volume} {82}},\ \bibinfo {pages} {035409} (\bibinfo {year} {2010})}\BibitemShut {NoStop}%
\bibitem [{\citenamefont {Lui}\ \emph {et~al.}(2011)\citenamefont {Lui}, \citenamefont {Li}, \citenamefont {Mak}, \citenamefont {Cappelluti},\ and\ \citenamefont {Heinz}}]{Lui2011}%
  \BibitemOpen
  \bibfield  {author} {\bibinfo {author} {\bibfnamefont {C.~H.}\ \bibnamefont {Lui}}, 
  \bibinfo {author} {\bibfnamefont {Z.}\ \bibnamefont {Li}}, 
  \bibinfo {author} {\bibfnamefont {K.~F.}\ \bibnamefont {Mak}}, 
  \bibinfo {author} {\bibfnamefont {E.}\ \bibnamefont {Cappelluti}},\ and\ 
  \bibinfo {author} {\bibfnamefont {T.~F.}\ \bibnamefont {Heinz}},\ }\bibfield  {title} 
  {\bibinfo {title} {Observation of an electrically tunable band gap in trilayer graphene},\ }
  \href {https://doi.org/10.1038/nphys2102} {\bibfield  {journal} {\bibinfo  {journal} {Nat. Phys.}\ }\textbf {\bibinfo {volume} {7}},\ \bibinfo {pages} {944--947} (\bibinfo {year} {2011})}\BibitemShut {NoStop}%
\bibitem [{\citenamefont {Jung}\ and\ \citenamefont {MacDonald}(2013)}]{Jung2013}%
  \BibitemOpen
  \bibfield  {author} {\bibinfo {author} {\bibfnamefont {J.}\ \bibnamefont {Jung}}\ and\ 
  \bibinfo {author} {\bibfnamefont {A.~H.}\ \bibnamefont {MacDonald}},\ }\bibfield  {title} 
  {\bibinfo {title} {Gapped broken symmetry states in {ABC}-stacked trilayer graphene},\ }
  \href {https://doi.org/10.1103/PhysRevB.88.075408} {\bibfield  {journal} {\bibinfo  {journal} {Phys. Rev. B}\ }\textbf {\bibinfo {volume} {88}},\ \bibinfo {pages} {075408} (\bibinfo {year} {2013})}\BibitemShut {NoStop}%
\bibitem [{\citenamefont {Jia}\ \emph {et~al.}(2013)\citenamefont {Jia}, \citenamefont {Gorbar},\ and\ \citenamefont {Gusynin}}]{Jia2013}%
  \BibitemOpen
  \bibfield  {author} {\bibinfo {author} {\bibfnamefont {J.}\ \bibnamefont {Jia}}, 
  \bibinfo {author} {\bibfnamefont {E.~V.}\ \bibnamefont {Gorbar}},\ and\ 
  \bibinfo {author} {\bibfnamefont {V.~P.}\ \bibnamefont {Gusynin}},\ }\bibfield  {title} 
  {\bibinfo {title} {Gap generation in {ABC}-stacked multilayer graphene: Screening versus band flattening},\ }
  \href {https://doi.org/10.1103/PhysRevB.88.205428} {\bibfield  {journal} {\bibinfo  {journal} {Phys. Rev. B}\ }\textbf {\bibinfo {volume} {88}},\ \bibinfo {pages} {205428} (\bibinfo {year} {2013})}\BibitemShut {NoStop}%
\bibitem [{\citenamefont {Lee}\ \emph {et~al.}(2014)\citenamefont {Lee}, \citenamefont {Tran}, \citenamefont {Myhro}, \citenamefont {Velasco}, \citenamefont {Gillgren}, \citenamefont {Lau}, \citenamefont {Barlas}, \citenamefont {Poumirol}, \citenamefont {Smirnov},\ and\ \citenamefont {Guinea}}]{Lee2014}%
  \BibitemOpen
  \bibfield  {author} {\bibinfo {author} {\bibfnamefont {Y.}\ \bibnamefont {Lee}}, 
  \bibinfo {author} {\bibfnamefont {D.}\ \bibnamefont {Tran}}, 
  \bibinfo {author} {\bibfnamefont {K.}\ \bibnamefont {Myhro}}, 
  \bibinfo {author} {\bibfnamefont {J.}\ \bibnamefont {Velasco}}, 
  \bibinfo {author} {\bibfnamefont {N.}\ \bibnamefont {Gillgren}}, 
  \bibinfo {author} {\bibfnamefont {C.~N.}\ \bibnamefont {Lau}}, 
  \bibinfo {author} {\bibfnamefont {Y.}\ \bibnamefont {Barlas}}, 
  \bibinfo {author} {\bibfnamefont {J.~M.}\ \bibnamefont {Poumirol}}, 
  \bibinfo {author} {\bibfnamefont {D.}\ \bibnamefont {Smirnov}},\ and\ 
  \bibinfo {author} {\bibfnamefont {F.}\ \bibnamefont {Guinea}},\ }\bibfield  {title} 
  {\bibinfo {title} {Competition between spontaneous symmetry breaking and single-particle gaps in trilayer graphene},\ }
  \href {https://doi.org/10.1038/ncomms6656} {\bibfield  {journal} {\bibinfo  {journal} {Nat. Commun.}\ }\textbf {\bibinfo {volume} {5}},\ \bibinfo {pages} {5656} (\bibinfo {year} {2014})}\BibitemShut {NoStop}%
\bibitem [{\citenamefont {Shi}\ \emph {et~al.}(2020)\citenamefont {Shi}, \citenamefont {Xu}, \citenamefont {Yang}, \citenamefont {Slizovskiy}, \citenamefont {Morozov}, \citenamefont {Son}, \citenamefont {Ozdemir}, \citenamefont {Mullan}, \citenamefont {Barrier}, \citenamefont {Yin}, \citenamefont {Berdyugin}, \citenamefont {Piot}, \citenamefont {Taniguchi}, \citenamefont {Watanabe}, \citenamefont {Fal'ko}, \citenamefont {Novoselov}, \citenamefont {Geim},\ and\ \citenamefont {Mishchenko}}]{Shi2020}%
  \BibitemOpen
  \bibfield  {author} {\bibinfo {author} {\bibfnamefont {Y.}\ \bibnamefont {Shi}}, 
  \bibinfo {author} {\bibfnamefont {S.}\ \bibnamefont {Xu}}, 
  \bibinfo {author} {\bibfnamefont {Y.}\ \bibnamefont {Yang}}, 
  \bibinfo {author} {\bibfnamefont {S.}\ \bibnamefont {Slizovskiy}}, 
  \bibinfo {author} {\bibfnamefont {S.~V.}\ \bibnamefont {Morozov}}, 
  \bibinfo {author} {\bibfnamefont {S.-K.}\ \bibnamefont {Son}}, 
  \bibinfo {author} {\bibfnamefont {S.}\ \bibnamefont {Ozdemir}}, 
  \bibinfo {author} {\bibfnamefont {C.}\ \bibnamefont {Mullan}}, 
  \bibinfo {author} {\bibfnamefont {J.}\ \bibnamefont {Barrier}}, 
  \bibinfo {author} {\bibfnamefont {J.}\ \bibnamefont {Yin}}, 
  \bibinfo {author} {\bibfnamefont {A.~I.}\ \bibnamefont {Berdyugin}}, 
  \bibinfo {author} {\bibfnamefont {B.~A.}\ \bibnamefont {Piot}}, 
  \bibinfo {author} {\bibfnamefont {T.}\ \bibnamefont {Taniguchi}}, 
  \bibinfo {author} {\bibfnamefont {K.}\ \bibnamefont {Watanabe}}, 
  \bibinfo {author} {\bibfnamefont {V.~I.}\ \bibnamefont {Fal'ko}}, 
  \bibinfo {author} {\bibfnamefont {K.~S.}\ \bibnamefont {Novoselov}}, 
  \bibinfo {author} {\bibfnamefont {A.~K.}\ \bibnamefont {Geim}},\ and\ 
  \bibinfo {author} {\bibfnamefont {A.}\ \bibnamefont {Mishchenko}},\ }\bibfield  {title} 
  {\bibinfo {title} {Electronic phase separation in multilayer rhombohedral graphite},\ }
  \href {https://doi.org/10.1038/s41586-020-2568-2} {\bibfield  {journal} {\bibinfo  {journal} {Nature}\ }\textbf {\bibinfo {volume} {584}},\ \bibinfo {pages} {210--214} (\bibinfo {year} {2020})}\BibitemShut {NoStop}%
\bibitem [{\citenamefont {Zhou}\ \emph {et~al.}(2021{\natexlab{a}})\citenamefont {Zhou}, \citenamefont {Xie}, \citenamefont {Ghazaryan}, \citenamefont {Holder}, \citenamefont {Ehrets}, \citenamefont {Spanton}, \citenamefont {Taniguchi}, \citenamefont {Watanabe}, \citenamefont {Berg}, \citenamefont {Serbyn},\ and\ \citenamefont {Young}}]{Zhou2021HQM}%
  \BibitemOpen
  \bibfield  {author} {\bibinfo {author} {\bibfnamefont {H.}\ \bibnamefont {Zhou}}, 
  \bibinfo {author} {\bibfnamefont {T.}\ \bibnamefont {Xie}}, 
  \bibinfo {author} {\bibfnamefont {A.}\ \bibnamefont {Ghazaryan}}, 
  \bibinfo {author} {\bibfnamefont {T.}\ \bibnamefont {Holder}}, 
  \bibinfo {author} {\bibfnamefont {J.~R.}\ \bibnamefont {Ehrets}}, 
  \bibinfo {author} {\bibfnamefont {E.~M.}\ \bibnamefont {Spanton}}, 
  \bibinfo {author} {\bibfnamefont {T.}\ \bibnamefont {Taniguchi}}, 
  \bibinfo {author} {\bibfnamefont {K.}\ \bibnamefont {Watanabe}}, 
  \bibinfo {author} {\bibfnamefont {E.}\ \bibnamefont {Berg}}, 
  \bibinfo {author} {\bibfnamefont {M.}\ \bibnamefont {Serbyn}},\ and\ 
  \bibinfo {author} {\bibfnamefont {A.~F.}\ \bibnamefont {Young}},\ }\bibfield  {title} 
  {\bibinfo {title} {Half- and quarter-metals in rhombohedral trilayer graphene},\ }
  \href {https://doi.org/10.1038/s41586-021-03938-w} {\bibfield  {journal} {\bibinfo  {journal} {Nature}\ }\textbf {\bibinfo {volume} {598}},\ \bibinfo {pages} {429--433} (\bibinfo {year} {2021}{\natexlab{a}})}\BibitemShut {NoStop}%
\bibitem [{\citenamefont {Zhou}\ \emph {et~al.}(2021{\natexlab{b}})\citenamefont {Zhou}, \citenamefont {Xie}, \citenamefont {Taniguchi}, \citenamefont {Watanabe},\ and\ \citenamefont {Young}}]{Zhou2021SC}%
  \BibitemOpen
  \bibfield  {author} {\bibinfo {author} {\bibfnamefont {H.}\ \bibnamefont {Zhou}}, 
  \bibinfo {author} {\bibfnamefont {T.}\ \bibnamefont {Xie}}, 
  \bibinfo {author} {\bibfnamefont {T.}\ \bibnamefont {Taniguchi}}, 
  \bibinfo {author} {\bibfnamefont {K.}\ \bibnamefont {Watanabe}},\ and\ 
  \bibinfo {author} {\bibfnamefont {A.~F.}\ \bibnamefont {Young}},\ }\bibfield  {title} 
  {\bibinfo {title} {Superconductivity in rhombohedral trilayer graphene},\ }
  \href {https://doi.org/10.1038/s41586-021-03926-0} {\bibfield  {journal} {\bibinfo  {journal} {Nature}\ }\textbf {\bibinfo {volume} {598}},\ \bibinfo {pages} {434--438} (\bibinfo {year} {2021}{\natexlab{b}})}\BibitemShut {NoStop}%
\bibitem [{\citenamefont {Winterer}\ \emph {et~al.}(2024)\citenamefont {Winterer}, \citenamefont {Geisenhof}, \citenamefont {Fernandez}, \citenamefont {Seiler}, \citenamefont {Zhang},\ and\ \citenamefont {Weitz}}]{Winterer2024}%
  \BibitemOpen
  \bibfield  {author} {\bibinfo {author} {\bibfnamefont {F.}\ \bibnamefont {Winterer}}, 
  \bibinfo {author} {\bibfnamefont {F.~R.}\ \bibnamefont {Geisenhof}}, 
  \bibinfo {author} {\bibfnamefont {N.}\ \bibnamefont {Fernandez}}, 
  \bibinfo {author} {\bibfnamefont {A.~M.}\ \bibnamefont {Seiler}}, 
  \bibinfo {author} {\bibfnamefont {F.}\ \bibnamefont {Zhang}},\ and\ 
  \bibinfo {author} {\bibfnamefont {R.~T.}\ \bibnamefont {Weitz}},\ }\bibfield  {title} 
  {\bibinfo {title} {Ferroelectric and spontaneous quantum Hall states in intrinsic rhombohedral trilayer graphene},\ }
  \href {https://doi.org/10.1038/s41567-023-02327-6} {\bibfield  {journal} {\bibinfo  {journal} {Nat. Phys.}\ }\textbf {\bibinfo {volume} {20}},\ \bibinfo {pages} {422--427} (\bibinfo {year} {2024})}\BibitemShut {NoStop}%
\bibitem [{\citenamefont {Cao}\ \emph {et~al.}(2018)\citenamefont {Cao}, \citenamefont {Fatemi}, \citenamefont {Demir}, \citenamefont {Fang}, \citenamefont {Tomarken}, \citenamefont {Luo}, \citenamefont {Sanchez-Yamagishi}, \citenamefont {Watanabe}, \citenamefont {Taniguchi}, \citenamefont {Kaxiras}, \citenamefont {Ashoori},\ and\ \citenamefont {Jarillo-Herrero}}]{Cao2018Ins}%
  \BibitemOpen
  \bibfield  {author} {\bibinfo {author} {\bibfnamefont {Y.}\ \bibnamefont {Cao}},
  \bibinfo {author} {\bibfnamefont {V.}\ \bibnamefont {Fatemi}},
  \bibinfo {author} {\bibfnamefont {A.}\ \bibnamefont {Demir}},
  \bibinfo {author} {\bibfnamefont {S.}\ \bibnamefont {Fang}},
  \bibinfo {author} {\bibfnamefont {S.~L.}\ \bibnamefont {Tomarken}},
  \bibinfo {author} {\bibfnamefont {J.~Y.}\ \bibnamefont {Luo}},
  \bibinfo {author} {\bibfnamefont {J.~D.}\ \bibnamefont {Sanchez-Yamagishi}},
  \bibinfo {author} {\bibfnamefont {K.}\ \bibnamefont {Watanabe}},
  \bibinfo {author} {\bibfnamefont {T.}\ \bibnamefont {Taniguchi}},
  \bibinfo {author} {\bibfnamefont {E.}\ \bibnamefont {Kaxiras}},
  \bibinfo {author} {\bibfnamefont {R.~C.}\ \bibnamefont {Ashoori}},\ and\
  \bibinfo {author} {\bibfnamefont {P.}\ \bibnamefont {Jarillo-Herrero}},\ }\bibfield  {title}
  {\bibinfo {title} {Correlated insulator behaviour at half-filling in magic-angle graphene superlattices},\ }
  \href {https://doi.org/10.1038/nature26154} {\bibfield  {journal} {\bibinfo  {journal} {Nature}\ }\textbf {\bibinfo {volume} {556}},\ \bibinfo {pages} {80--84} (\bibinfo {year} {2018})}\BibitemShut {NoStop}%
\bibitem [{\citenamefont {Cao}\ \emph {et~al.}(2018)\citenamefont {Cao}, \citenamefont {Fatemi}, \citenamefont {Fang}, \citenamefont {Watanabe}, \citenamefont {Taniguchi}, \citenamefont {Kaxiras},\ and\ \citenamefont {Jarillo-Herrero}}]{Cao2018SC}%
  \BibitemOpen
  \bibfield  {author} {\bibinfo {author} {\bibfnamefont {Y.}\ \bibnamefont {Cao}},
  \bibinfo {author} {\bibfnamefont {V.}\ \bibnamefont {Fatemi}},
  \bibinfo {author} {\bibfnamefont {S.}\ \bibnamefont {Fang}},
  \bibinfo {author} {\bibfnamefont {K.}\ \bibnamefont {Watanabe}},
  \bibinfo {author} {\bibfnamefont {T.}\ \bibnamefont {Taniguchi}},
  \bibinfo {author} {\bibfnamefont {E.}\ \bibnamefont {Kaxiras}},\ and\
  \bibinfo {author} {\bibfnamefont {P.}\ \bibnamefont {Jarillo-Herrero}},\ }\bibfield  {title}
  {\bibinfo {title} {Unconventional superconductivity in magic-angle graphene superlattices},\ }
  \href {https://doi.org/10.1038/nature26160} {\bibfield  {journal} {\bibinfo  {journal} {Nature}\ }\textbf {\bibinfo {volume} {556}},\ \bibinfo {pages} {43--50} (\bibinfo {year} {2018})}\BibitemShut {NoStop}%
\bibitem [{\citenamefont {Sharpe}\ \emph {et~al.}(2019)\citenamefont {Sharpe}, \citenamefont {Fox}, \citenamefont {Barnard}, \citenamefont {Finney}, \citenamefont {Watanabe}, \citenamefont {Taniguchi}, \citenamefont {Kastner},\ and\ \citenamefont {Goldhaber-Gordon}}]{Sharpe2019}%
  \BibitemOpen
  \bibfield  {author} {\bibinfo {author} {\bibfnamefont {A.~L.}\ \bibnamefont {Sharpe}},
  \bibinfo {author} {\bibfnamefont {E.~J.}\ \bibnamefont {Fox}},
  \bibinfo {author} {\bibfnamefont {A.~W.}\ \bibnamefont {Barnard}},
  \bibinfo {author} {\bibfnamefont {N.}\ \bibnamefont {Finney}},
  \bibinfo {author} {\bibfnamefont {K.}\ \bibnamefont {Watanabe}},
  \bibinfo {author} {\bibfnamefont {T.}\ \bibnamefont {Taniguchi}},
  \bibinfo {author} {\bibfnamefont {M.~A.}\ \bibnamefont {Kastner}},\ and\
  \bibinfo {author} {\bibfnamefont {D.}\ \bibnamefont {Goldhaber-Gordon}},\ }\bibfield  {title}
  {\bibinfo {title} {Emergent ferromagnetism near three-quarters filling in twisted bilayer graphene},\ }
  \href {https://doi.org/10.1126/science.aaw3780} {\bibfield  {journal} {\bibinfo  {journal} {Science}\ }\textbf {\bibinfo {volume} {365}},\ \bibinfo {pages} {605--608} (\bibinfo {year} {2019})}\BibitemShut {NoStop}%
\bibitem [{\citenamefont {Serlin}\ \emph {et~al.}(2020)\citenamefont {Serlin}, \citenamefont {Tschirhart}, \citenamefont {Polshyn}, \citenamefont {Zhang}, \citenamefont {Zhao}, \citenamefont {Watanabe}, \citenamefont {Taniguchi}, \citenamefont {Balents},\ and\ \citenamefont {Young}}]{Serlin2020}%
  \BibitemOpen
  \bibfield  {author} {\bibinfo {author} {\bibfnamefont {M.}\ \bibnamefont {Serlin}},
  \bibinfo {author} {\bibfnamefont {C.~L.}\ \bibnamefont {Tschirhart}},
  \bibinfo {author} {\bibfnamefont {H.}\ \bibnamefont {Polshyn}},
  \bibinfo {author} {\bibfnamefont {Y.}\ \bibnamefont {Zhang}},
  \bibinfo {author} {\bibfnamefont {J.}\ \bibnamefont {Zhao}},
  \bibinfo {author} {\bibfnamefont {K.}\ \bibnamefont {Watanabe}},
  \bibinfo {author} {\bibfnamefont {T.}\ \bibnamefont {Taniguchi}},
  \bibinfo {author} {\bibfnamefont {L.}\ \bibnamefont {Balents}},\ and\
  \bibinfo {author} {\bibfnamefont {A.~F.}\ \bibnamefont {Young}},\ }\bibfield  {title}
  {\bibinfo {title} {Intrinsic quantized anomalous Hall effect in a moir\'e heterostructure},\ }
  \href {https://doi.org/10.1126/science.aay5533} {\bibfield  {journal} {\bibinfo  {journal} {Science}\ }\textbf {\bibinfo {volume} {367}},\ \bibinfo {pages} {900--903} (\bibinfo {year} {2020})}\BibitemShut {NoStop}%
\bibitem [{\citenamefont {Xie}\ \emph {et~al.}(2021)\citenamefont {Xie}, \citenamefont {Pierce}, \citenamefont {Park}, \citenamefont {Parker}, \citenamefont {Khalaf}, \citenamefont {Ledwith}, \citenamefont {Cao}, \citenamefont {Lee}, \citenamefont {Chen}, \citenamefont {Forrester}, \citenamefont {Watanabe}, \citenamefont {Taniguchi}, \citenamefont {Vishwanath}, \citenamefont {Jarillo-Herrero},\ and\ \citenamefont {Yacoby}}]{Xie2021FCI}%
  \BibitemOpen
  \bibfield  {author} {\bibinfo {author} {\bibfnamefont {Y.}\ \bibnamefont {Xie}},
  \bibinfo {author} {\bibfnamefont {A.~T.}\ \bibnamefont {Pierce}},
  \bibinfo {author} {\bibfnamefont {J.~M.}\ \bibnamefont {Park}},
  \bibinfo {author} {\bibfnamefont {D.~E.}\ \bibnamefont {Parker}},
  \bibinfo {author} {\bibfnamefont {E.}\ \bibnamefont {Khalaf}},
  \bibinfo {author} {\bibfnamefont {P.~J.}\ \bibnamefont {Ledwith}},
  \bibinfo {author} {\bibfnamefont {Y.}\ \bibnamefont {Cao}},
  \bibinfo {author} {\bibfnamefont {S.~H.}\ \bibnamefont {Lee}},
  \bibinfo {author} {\bibfnamefont {S.}\ \bibnamefont {Chen}},
  \bibinfo {author} {\bibfnamefont {P.~R.}\ \bibnamefont {Forrester}},
  \bibinfo {author} {\bibfnamefont {K.}\ \bibnamefont {Watanabe}},
  \bibinfo {author} {\bibfnamefont {T.}\ \bibnamefont {Taniguchi}},
  \bibinfo {author} {\bibfnamefont {A.}\ \bibnamefont {Vishwanath}},
  \bibinfo {author} {\bibfnamefont {P.}\ \bibnamefont {Jarillo-Herrero}},\ and\
  \bibinfo {author} {\bibfnamefont {A.}\ \bibnamefont {Yacoby}},\ }\bibfield  {title}
  {\bibinfo {title} {Fractional Chern insulators in magic-angle twisted bilayer graphene},\ }
  \href {https://doi.org/10.1038/s41586-021-04142-6} {\bibfield  {journal} {\bibinfo  {journal} {Nature}\ }\textbf {\bibinfo {volume} {600}},\ \bibinfo {pages} {439--443} (\bibinfo {year} {2021})}\BibitemShut {NoStop}%
\bibitem [{\citenamefont {Cao}\ \emph {et~al.}(2021)\citenamefont {Cao}, \citenamefont {Rodan-Legrain}, \citenamefont {Park}, \citenamefont {Yuan}, \citenamefont {Watanabe}, \citenamefont {Taniguchi}, \citenamefont {Fernandes}, \citenamefont {Fu},\ and\ \citenamefont {Jarillo-Herrero}}]{Cao2021Nematic}%
  \BibitemOpen
  \bibfield  {author} {\bibinfo {author} {\bibfnamefont {Y.}\ \bibnamefont {Cao}},
  \bibinfo {author} {\bibfnamefont {D.}\ \bibnamefont {Rodan-Legrain}},
  \bibinfo {author} {\bibfnamefont {J.~M.}\ \bibnamefont {Park}},
  \bibinfo {author} {\bibfnamefont {N.~F.~Q.}\ \bibnamefont {Yuan}},
  \bibinfo {author} {\bibfnamefont {K.}\ \bibnamefont {Watanabe}},
  \bibinfo {author} {\bibfnamefont {T.}\ \bibnamefont {Taniguchi}},
  \bibinfo {author} {\bibfnamefont {R.~M.}\ \bibnamefont {Fernandes}},
  \bibinfo {author} {\bibfnamefont {L.}\ \bibnamefont {Fu}},\ and\
  \bibinfo {author} {\bibfnamefont {P.}\ \bibnamefont {Jarillo-Herrero}},\ }\bibfield  {title}
  {\bibinfo {title} {Nematicity and competing orders in superconducting magic-angle graphene},\ }
  \href {https://doi.org/10.1126/science.abc2836} {\bibfield  {journal} {\bibinfo  {journal} {Science}\ }\textbf {\bibinfo {volume} {372}},\ \bibinfo {pages} {264--271} (\bibinfo {year} {2021})}\BibitemShut {NoStop}%
\bibitem [{\citenamefont {Park}\ \emph {et~al.}(2021)\citenamefont {Park}, \citenamefont {Cao}, \citenamefont {Watanabe}, \citenamefont {Taniguchi},\ and\ \citenamefont {Jarillo-Herrero}}]{Park2021TTG}%
  \BibitemOpen
  \bibfield  {author} {\bibinfo {author} {\bibfnamefont {J.~M.}\ \bibnamefont {Park}},
  \bibinfo {author} {\bibfnamefont {Y.}\ \bibnamefont {Cao}},
  \bibinfo {author} {\bibfnamefont {K.}\ \bibnamefont {Watanabe}},
  \bibinfo {author} {\bibfnamefont {T.}\ \bibnamefont {Taniguchi}},\ and\
  \bibinfo {author} {\bibfnamefont {P.}\ \bibnamefont {Jarillo-Herrero}},\ }\bibfield  {title}
  {\bibinfo {title} {Tunable strongly coupled superconductivity in magic-angle twisted trilayer graphene},\ }
  \href {https://doi.org/10.1038/s41586-021-03192-0} {\bibfield  {journal} {\bibinfo  {journal} {Nature}\ }\textbf {\bibinfo {volume} {590}},\ \bibinfo {pages} {249--255} (\bibinfo {year} {2021})}\BibitemShut {NoStop}%
\bibitem [{\citenamefont {Hao}\ \emph {et~al.}(2021)\citenamefont {Hao}, \citenamefont {Zimmerman}, \citenamefont {Ledwith}, \citenamefont {Khalaf}, \citenamefont {Haei Najafabadi}, \citenamefont {Watanabe}, \citenamefont {Taniguchi}, \citenamefont {Vishwanath},\ and\ \citenamefont {Kim}}]{Hao2021ATTG}%
  \BibitemOpen
  \bibfield  {author} {\bibinfo {author} {\bibfnamefont {Z.}\ \bibnamefont {Hao}},
  \bibinfo {author} {\bibfnamefont {A.~M.}\ \bibnamefont {Zimmerman}},
  \bibinfo {author} {\bibfnamefont {P.}\ \bibnamefont {Ledwith}},
  \bibinfo {author} {\bibfnamefont {E.}\ \bibnamefont {Khalaf}},
  \bibinfo {author} {\bibfnamefont {D.}\ \bibnamefont {Haei Najafabadi}},
  \bibinfo {author} {\bibfnamefont {K.}\ \bibnamefont {Watanabe}},
  \bibinfo {author} {\bibfnamefont {T.}\ \bibnamefont {Taniguchi}},
  \bibinfo {author} {\bibfnamefont {A.}\ \bibnamefont {Vishwanath}},\ and\
  \bibinfo {author} {\bibfnamefont {P.}\ \bibnamefont {Kim}},\ }\bibfield  {title}
  {\bibinfo {title} {Electric field--tunable superconductivity in alternating-twist magic-angle trilayer graphene},\ }
  \href {https://doi.org/10.1126/science.abg0399} {\bibfield  {journal} {\bibinfo  {journal} {Science}\ }\textbf {\bibinfo {volume} {371}},\ \bibinfo {pages} {1133--1138} (\bibinfo {year} {2021})}\BibitemShut {NoStop}%
\bibitem [{\citenamefont {Park}\ \emph {et~al.}(2022)\citenamefont {Park}, \citenamefont {Cao}, \citenamefont {Xia}, \citenamefont {Sun}, \citenamefont {Watanabe}, \citenamefont {Taniguchi},\ and\ \citenamefont {Jarillo-Herrero}}]{Park2022MAMLG}%
  \BibitemOpen
  \bibfield  {author} {\bibinfo {author} {\bibfnamefont {J.~M.}\ \bibnamefont {Park}},
  \bibinfo {author} {\bibfnamefont {Y.}\ \bibnamefont {Cao}},
  \bibinfo {author} {\bibfnamefont {L.-Q.}\ \bibnamefont {Xia}},
  \bibinfo {author} {\bibfnamefont {S.}\ \bibnamefont {Sun}},
  \bibinfo {author} {\bibfnamefont {K.}\ \bibnamefont {Watanabe}},
  \bibinfo {author} {\bibfnamefont {T.}\ \bibnamefont {Taniguchi}},\ and\
  \bibinfo {author} {\bibfnamefont {P.}\ \bibnamefont {Jarillo-Herrero}},\ }\bibfield  {title}
  {\bibinfo {title} {Robust superconductivity in magic-angle multilayer graphene family},\ }
  \href {https://doi.org/10.1038/s41563-022-01287-1} {\bibfield  {journal} {\bibinfo  {journal} {Nat. Mater.}\ }\textbf {\bibinfo {volume} {21}},\ \bibinfo {pages} {877--883} (\bibinfo {year} {2022})}\BibitemShut {NoStop}%
\bibitem [{\citenamefont {Burg}\ \emph {et~al.}(2022)\citenamefont {Burg}, \citenamefont {Khalaf}, \citenamefont {Wang}, \citenamefont {Watanabe}, \citenamefont {Taniguchi},\ and\ \citenamefont {Tutuc}}]{Burg2022ATQG}%
  \BibitemOpen
  \bibfield  {author} {\bibinfo {author} {\bibfnamefont {G.~W.}\ \bibnamefont {Burg}},
  \bibinfo {author} {\bibfnamefont {E.}\ \bibnamefont {Khalaf}},
  \bibinfo {author} {\bibfnamefont {Y.}\ \bibnamefont {Wang}},
  \bibinfo {author} {\bibfnamefont {K.}\ \bibnamefont {Watanabe}},
  \bibinfo {author} {\bibfnamefont {T.}\ \bibnamefont {Taniguchi}},\ and\
  \bibinfo {author} {\bibfnamefont {E.}\ \bibnamefont {Tutuc}},\ }\bibfield  {title}
  {\bibinfo {title} {Emergence of correlations in alternating twist quadrilayer graphene},\ }
  \href {https://doi.org/10.1038/s41563-022-01286-2} {\bibfield  {journal} {\bibinfo  {journal} {Nat. Mater.}\ }\textbf {\bibinfo {volume} {21}},\ \bibinfo {pages} {884--889} (\bibinfo {year} {2022})}\BibitemShut {NoStop}%
\bibitem [{\citenamefont {Kim}\ \emph {et~al.}(2013)\citenamefont {Kim}, \citenamefont {Yun}, \citenamefont {Nam}, \citenamefont {Son}, \citenamefont {Lee}, \citenamefont {Kim}, \citenamefont {Seo}, \citenamefont {Choi}, \citenamefont {Lee}, \citenamefont {Lee},\ and\ \citenamefont {Kim}}]{Kim2019}%
  \BibitemOpen
  \bibfield  {author} {\bibinfo {author} {\bibfnamefont {Y.}~\bibnamefont {Kim}}, 
  \bibinfo {author} {\bibfnamefont {H.}~\bibnamefont {Yun}}, 
  \bibinfo {author} {\bibfnamefont {S.-G.}\ \bibnamefont {Nam}}, 
  \bibinfo {author} {\bibfnamefont {M.}~\bibnamefont {Son}}, 
  \bibinfo {author} {\bibfnamefont {D.~S.}\ \bibnamefont {Lee}}, 
  \bibinfo {author} {\bibfnamefont {D.~C.}\ \bibnamefont {Kim}}, 
  \bibinfo {author} {\bibfnamefont {S.}~\bibnamefont {Seo}}, 
  \bibinfo {author} {\bibfnamefont {H.~C.}\ \bibnamefont {Choi}}, 
  \bibinfo {author} {\bibfnamefont {H.-J.}\ \bibnamefont {Lee}}, 
  \bibinfo {author} {\bibfnamefont {S.~W.}\ \bibnamefont {Lee}},\ and\ 
  \bibinfo {author} {\bibfnamefont {J.~S.}\ \bibnamefont {Kim}},\ }\bibfield  {title} 
  {\bibinfo {title} {Breakdown of the Interlayer Coherence in Twisted Bilayer Graphene},\ }
  \href {https://doi.org/10.1103/PhysRevLett.110.096602} {\bibfield  {journal} {\bibinfo  {journal} {Phys. Rev. Lett.}\ }\textbf {\bibinfo {volume} {110}},\ \bibinfo {pages} {096602} (\bibinfo {year} {2013})}\BibitemShut {NoStop}%
\bibitem [{\citenamefont {Kim}\ \emph {et~al.}(2021)\citenamefont {Kim}, \citenamefont {Moon}, \citenamefont {Watanabe}, \citenamefont {Taniguchi},\ and\ \citenamefont {Smet}}]{Kim2021}%
  \BibitemOpen
  \bibfield  {author} {\bibinfo {author} {\bibfnamefont {Y.}~\bibnamefont {Kim}}, 
  \bibinfo {author} {\bibfnamefont {P.}~\bibnamefont {Moon}}, 
  \bibinfo {author} {\bibfnamefont {K.}~\bibnamefont {Watanabe}}, 
  \bibinfo {author} {\bibfnamefont {T.}~\bibnamefont {Taniguchi}},\ and\ 
  \bibinfo {author} {\bibfnamefont {J.~H.}\ \bibnamefont {Smet}},\ }\bibfield  {title} 
  {\bibinfo {title} {Odd Integer Quantum Hall States with Interlayer Coherence in Twisted Bilayer Graphene},\ }
  \href {https://doi.org/10.1021/acs.nanolett.1c00360} {\bibfield  {journal} {\bibinfo  {journal} {Nano Lett.}\ }\textbf {\bibinfo {volume} {21}},\ \bibinfo {pages} {4249--4254} (\bibinfo {year} {2021})}\BibitemShut {NoStop}%
\bibitem [{\citenamefont {Shi}\ \emph {et~al.}(2022)\citenamefont {Shi}, \citenamefont {Shih}, \citenamefont {Rhodes}, \citenamefont {Kim}, \citenamefont {Barmak}, \citenamefont {Watanabe}, \citenamefont {Taniguchi}, \citenamefont {Papi\'c}, \citenamefont {Abanin}, \citenamefont {Hone},\ and\ \citenamefont {Dean}}]{Shi2022WSe2EC}%
  \BibitemOpen
  \bibfield  {author} {\bibinfo {author} {\bibfnamefont {Q.}\ \bibnamefont {Shi}}, 
  \bibinfo {author} {\bibfnamefont {E.-M.}\ \bibnamefont {Shih}}, 
  \bibinfo {author} {\bibfnamefont {D.}\ \bibnamefont {Rhodes}}, 
  \bibinfo {author} {\bibfnamefont {B.}\ \bibnamefont {Kim}}, 
  \bibinfo {author} {\bibfnamefont {K.}\ \bibnamefont {Barmak}}, 
  \bibinfo {author} {\bibfnamefont {K.}\ \bibnamefont {Watanabe}}, 
  \bibinfo {author} {\bibfnamefont {T.}\ \bibnamefont {Taniguchi}}, 
  \bibinfo {author} {\bibfnamefont {Z.}\ \bibnamefont {Papi\'c}}, 
  \bibinfo {author} {\bibfnamefont {D.~A.}\ \bibnamefont {Abanin}}, 
  \bibinfo {author} {\bibfnamefont {J.}\ \bibnamefont {Hone}},\ and\ 
  \bibinfo {author} {\bibfnamefont {C.~R.}\ \bibnamefont {Dean}},\ }\bibfield  {title} 
  {\bibinfo {title} {Bilayer WSe$_2$ as a natural platform for interlayer exciton condensates in the strong coupling limit},\ }
  \href {https://doi.org/10.1038/s41565-022-01104-5} {\bibfield  {journal} {\bibinfo  {journal} {Nat. Nanotechnol.}\ }\textbf {\bibinfo {volume} {17}},\ \bibinfo {pages} {577--582} (\bibinfo {year} {2022})}\BibitemShut {NoStop}%
\bibitem [{\citenamefont {Kim}\ \emph {et~al.}(2023)\citenamefont {Kim}, \citenamefont {Kang}, \citenamefont {Choi}, \citenamefont {Watanabe}, \citenamefont {Taniguchi}, \citenamefont {Lee}, \citenamefont {Cho},\ and\ \citenamefont {Kim}}]{Kim2023}%
  \BibitemOpen
  \bibfield  {author} {\bibinfo {author} {\bibfnamefont {D.}~\bibnamefont {Kim}}, 
  \bibinfo {author} {\bibfnamefont {B.}~\bibnamefont {Kang}}, 
  \bibinfo {author} {\bibfnamefont {Y.-B.}\ \bibnamefont {Choi}}, 
  \bibinfo {author} {\bibfnamefont {K.}~\bibnamefont {Watanabe}}, 
  \bibinfo {author} {\bibfnamefont {T.}~\bibnamefont {Taniguchi}}, 
  \bibinfo {author} {\bibfnamefont {G.-H.}\ \bibnamefont {Lee}}, 
  \bibinfo {author} {\bibfnamefont {G.~Y.}\ \bibnamefont {Cho}},\ and\ 
  \bibinfo {author} {\bibfnamefont {Y.}~\bibnamefont {Kim}},\ }\bibfield  {title} 
  {\bibinfo {title} {Robust interlayer-coherent quantum Hall states in twisted bilayer graphene},\ }
  \href {https://doi.org/10.1021/acs.nanolett.2c03836} {\bibfield  {journal} {\bibinfo  {journal} {Nano Lett.}\ }\textbf {\bibinfo {volume} {23}},\ \bibinfo {pages} {163--169} (\bibinfo {year} {2023})}\BibitemShut {NoStop}%
\bibitem [{\citenamefont {Kim}\ \emph {et~al.}(2023)\citenamefont {Kim}, \citenamefont {Kim}, \citenamefont {Watanabe}, \citenamefont {Taniguchi}, \citenamefont {Smet},\ and\ \citenamefont {Kim}}]{Kim2023AdvSci}%
  \BibitemOpen
  \bibfield  {author} {\bibinfo {author} {\bibfnamefont {S.}~\bibnamefont {Kim}}, 
  \bibinfo {author} {\bibfnamefont {D.}~\bibnamefont {Kim}}, 
  \bibinfo {author} {\bibfnamefont {K.}~\bibnamefont {Watanabe}}, 
  \bibinfo {author} {\bibfnamefont {T.}~\bibnamefont {Taniguchi}}, 
  \bibinfo {author} {\bibfnamefont {J.~H.}\ \bibnamefont {Smet}},\ and\ 
  \bibinfo {author} {\bibfnamefont {Y.}~\bibnamefont {Kim}},\ }\bibfield  {title} 
  {\bibinfo {title} {Orbitally controlled quantum Hall states in decoupled two-bilayer graphene sheets},\ }
  \href {https://doi.org/10.1002/advs.202300574} {\bibfield  {journal} {\bibinfo  {journal} {Adv. Sci.}\ }\textbf {\bibinfo {volume} {10}},\ \bibinfo {pages} {2300574} (\bibinfo {year} {2023})}\BibitemShut {NoStop}%
\bibitem [{\citenamefont {Kim}\ \emph {et~al.}(2025)\citenamefont {Kim}, \citenamefont {Jin}, \citenamefont {Taniguchi}, \citenamefont {Watanabe}, \citenamefont {Smet}, \citenamefont {Cho},\ and\ \citenamefont {Kim}}]{Kim2025}%
  \BibitemOpen
  \bibfield  {author} {\bibinfo {author} {\bibfnamefont {D.}~\bibnamefont {Kim}}, 
  \bibinfo {author} {\bibfnamefont {S.}~\bibnamefont {Jin}}, 
  \bibinfo {author} {\bibfnamefont {T.}~\bibnamefont {Taniguchi}}, 
  \bibinfo {author} {\bibfnamefont {K.}~\bibnamefont {Watanabe}}, 
  \bibinfo {author} {\bibfnamefont {J.~H.}\ \bibnamefont {Smet}}, 
  \bibinfo {author} {\bibfnamefont {G.~Y.}\ \bibnamefont {Cho}},\ and\ 
  \bibinfo {author} {\bibfnamefont {Y.}~\bibnamefont {Kim}},\ }\bibfield  {title} 
  {\bibinfo {title} {Observation of 1/3 fractional quantum Hall physics in balanced large angle twisted bilayer graphene},\ }
  \href {https://doi.org/10.1038/s41467-024-55486-2} {\bibfield  {journal} {\bibinfo  {journal} {Nat. Commun.}\ }\textbf {\bibinfo {volume} {16}},\ \bibinfo {pages} {179} (\bibinfo {year} {2025})}\BibitemShut {NoStop}%
\bibitem [{\citenamefont {Li}\ \emph {et~al.}(2024)\citenamefont {Li}, \citenamefont {Chen}, \citenamefont {Wei}, \citenamefont {Chen}, \citenamefont {Huang}, \citenamefont {Zhu}, \citenamefont {Zhu}, \citenamefont {An}, \citenamefont {Song}, \citenamefont {Gan}, \citenamefont {Zhang}, \citenamefont {Watanabe}, \citenamefont {Taniguchi}, \citenamefont {Shi}, \citenamefont {Novoselov}, \citenamefont {Wang}, \citenamefont {Yu},\ and\ \citenamefont {Wang}}]{Li2024}%
  \BibitemOpen
  \bibfield  {author} {\bibinfo {author} {\bibfnamefont {Q.}~\bibnamefont {Li}}, 
  \bibinfo {author} {\bibfnamefont {Y.}~\bibnamefont {Chen}}, 
  \bibinfo {author} {\bibfnamefont {L.}~\bibnamefont {Wei}}, 
  \bibinfo {author} {\bibfnamefont {H.}~\bibnamefont {Chen}}, 
  \bibinfo {author} {\bibfnamefont {Y.}~\bibnamefont {Huang}}, 
  \bibinfo {author} {\bibfnamefont {Y.}~\bibnamefont {Zhu}}, 
  \bibinfo {author} {\bibfnamefont {W.}~\bibnamefont {Zhu}}, 
  \bibinfo {author} {\bibfnamefont {D.}~\bibnamefont {An}}, 
  \bibinfo {author} {\bibfnamefont {J.}~\bibnamefont {Song}}, 
  \bibinfo {author} {\bibfnamefont {Q.}~\bibnamefont {Gan}}, 
  \bibinfo {author} {\bibfnamefont {Q.}~\bibnamefont {Zhang}}, 
  \bibinfo {author} {\bibfnamefont {K.}~\bibnamefont {Watanabe}}, 
  \bibinfo {author} {\bibfnamefont {T.}~\bibnamefont {Taniguchi}}, 
  \bibinfo {author} {\bibfnamefont {X.}~\bibnamefont {Shi}}, 
  \bibinfo {author} {\bibfnamefont {K.~S.}\ \bibnamefont {Novoselov}}, 
  \bibinfo {author} {\bibfnamefont {R.}~\bibnamefont {Wang}}, 
  \bibinfo {author} {\bibfnamefont {G.}~\bibnamefont {Yu}},\ and\ 
  \bibinfo {author} {\bibfnamefont {L.}~\bibnamefont {Wang}},\ }\bibfield  {title} 
  {\bibinfo {title} {Strongly coupled magneto-exciton condensates in large-angle twisted double bilayer graphene},\ }
  \href {https://doi.org/10.1038/s41467-024-49406-7} {\bibfield  {journal} {\bibinfo  {journal} {Nat. Commun.}\ }\textbf {\bibinfo {volume} {15}},\ \bibinfo {pages} {5065} (\bibinfo {year} {2024})}\BibitemShut {NoStop}%
\bibitem [{\citenamefont {Davydov}\ \emph {et~al.}(2025)\citenamefont {Davydov}, \citenamefont {Long}, \citenamefont {Tavakley}, \citenamefont {Watanabe}, \citenamefont {Taniguchi},\ and\ \citenamefont {Wang}}]{Davydov2025}%
  \BibitemOpen
  \bibfield  {author} {\bibinfo {author} {\bibfnamefont {K.}~\bibnamefont {Davydov}}, 
  \bibinfo {author} {\bibfnamefont {D.}~\bibnamefont {Long}}, 
  \bibinfo {author} {\bibfnamefont {J.~A.}~\bibnamefont {Tavakley}}, 
  \bibinfo {author} {\bibfnamefont {K.}~\bibnamefont {Watanabe}}, 
  \bibinfo {author} {\bibfnamefont {T.}~\bibnamefont {Taniguchi}},\ and\ 
  \bibinfo {author} {\bibfnamefont {K.}~\bibnamefont {Wang}},\ }\bibfield  {title} 
  {\bibinfo {title} {Stability Diagram of Layer-Polarized Quantum Hall States in Twisted Trilayer Graphene},\ }
  \href {https://doi.org/10.1021/acs.jpclett.5c01221} {\bibfield  {journal} {\bibinfo  {journal} {J. Phys. Chem. Lett.}\ }\textbf {\bibinfo {volume} {16}},\ \bibinfo {pages} {7990--7997} (\bibinfo {year} {2025})}\BibitemShut {NoStop}%
\bibitem [{\citenamefont {Sanchez-Yamagishi}\ \emph {et~al.}(2017)\citenamefont {Sanchez-Yamagishi}, \citenamefont {Luo}, \citenamefont {Young}, \citenamefont {Hunt}, \citenamefont {Watanabe}, \citenamefont {Taniguchi}, \citenamefont {Ashoori},\ and\ \citenamefont {Jarillo-Herrero}}]{SanchezYamagishi2017}%
  \BibitemOpen
  \bibfield  {author} {\bibinfo {author} {\bibfnamefont {J.~D.}\ \bibnamefont {Sanchez-Yamagishi}}, 
  \bibinfo {author} {\bibfnamefont {J.~Y.}\ \bibnamefont {Luo}}, 
  \bibinfo {author} {\bibfnamefont {A.~F.}\ \bibnamefont {Young}}, 
  \bibinfo {author} {\bibfnamefont {B.~M.}\ \bibnamefont {Hunt}}, 
  \bibinfo {author} {\bibfnamefont {K.}~\bibnamefont {Watanabe}}, 
  \bibinfo {author} {\bibfnamefont {T.}~\bibnamefont {Taniguchi}}, 
  \bibinfo {author} {\bibfnamefont {R.~C.}\ \bibnamefont {Ashoori}},\ and\ 
  \bibinfo {author} {\bibfnamefont {P.}~\bibnamefont {Jarillo-Herrero}},\ }\bibfield  {title} 
  {\bibinfo {title} {Helical edge states and fractional quantum Hall effect in a graphene electron–hole bilayer},\ }
  \href {https://doi.org/10.1038/nnano.2016.214} {\bibfield  {journal} {\bibinfo  {journal} {Nat. Nanotechnol.}\ }\textbf {\bibinfo {volume} {12}},\ \bibinfo {pages} {118--122} (\bibinfo {year} {2017})}\BibitemShut {NoStop}%
\bibitem [{\citenamefont {Hoke}\ \emph {et~al.}(2024)\citenamefont {Hoke}, \citenamefont {Li}, \citenamefont {May-Mann}, \citenamefont {Watanabe}, \citenamefont {Taniguchi}, \citenamefont {Bradlyn}, \citenamefont {Hughes},\ and\ \citenamefont {Feldman}}]{Hoke2024}%
  \BibitemOpen
  \bibfield  {author} {\bibinfo {author} {\bibfnamefont {J.~C.}\ \bibnamefont {Hoke}}, 
  \bibinfo {author} {\bibfnamefont {Y.}~\bibnamefont {Li}}, 
  \bibinfo {author} {\bibfnamefont {J.}~\bibnamefont {May-Mann}}, 
  \bibinfo {author} {\bibfnamefont {K.}~\bibnamefont {Watanabe}}, 
  \bibinfo {author} {\bibfnamefont {T.}~\bibnamefont {Taniguchi}}, 
  \bibinfo {author} {\bibfnamefont {B.}~\bibnamefont {Bradlyn}}, 
  \bibinfo {author} {\bibfnamefont {T.~L.}\ \bibnamefont {Hughes}},\ and\ 
  \bibinfo {author} {\bibfnamefont {B.~E.}\ \bibnamefont {Feldman}},\ }\bibfield  {title} 
  {\bibinfo {title} {Uncovering the spin ordering in magic-angle graphene via edge state equilibration},\ }
  \href {https://doi.org/10.1038/s41467-024-48385-z} {\bibfield  {journal} {\bibinfo  {journal} {Nat. Commun.}\ }\textbf {\bibinfo {volume} {15}},\ \bibinfo {pages} {4321} (\bibinfo {year} {2024})}\BibitemShut {NoStop}%
\end{thebibliography}
\end{document}